\documentclass[a4paper,11pt]{article}
\pdfoutput=1
\usepackage{jheppub}
\usepackage{amsmath}
\usepackage{amssymb}
\usepackage{color}
\usepackage{csquotes}
\usepackage{graphicx}
\usepackage{hyperref}
\usepackage{xspace,slashed}
\usepackage[utf8]{inputenc}
\usepackage{multirow}
\usepackage{algorithm}
\usepackage{algorithmic}
\usepackage{stackengine}
\usepackage{enumitem}
\usepackage{etoolbox}
\usepackage{siunitx}
\usepackage{tensor}

\newtoggle{draft}

\togglefalse{draft}

\iftoggle{draft} {
  \usepackage{changes}
}{%
  \usepackage[final]{changes}
}

\definechangesauthor[name=pn, color=blue]{pn}
\definechangesauthor[name=km, color=magenta]{km}

\usepackage{booktabs}

\makeatletter
\g@addto@macro\bfseries{\boldmath}
\makeatother

\newcommand{\xa}{\mathfrak{m}}
\newcommand{\yb}{\mathfrak{n}}

\newcommand{\ep}{\epsilon}

\iftoggle{draft} {
  \newcommand{\tobedeleted}[1]{\textcolor{azure}{#1}}
  
}{%
  \newcommand{\tobedeleted}[1]{}
  
}

\newcommand{\be}{\begin{equation}}
\newcommand{\ee}{\end{equation}}

\usepackage{soul}
\allowdisplaybreaks

\definecolor{azure}{rgb}{0.0, 0.9, 1.0}

\newcommand{\TTPaff}{Institute for Theoretical Particle Physics,
  KIT, 76128 Karlsruhe, Germany}
\newcommand{\IAPaff}{Institute for Astroparticle Physics, KIT, 76344 Eggenstein-Leopoldshafen, Germany}

\preprint{
  \begin {flushright}
    TTP25-032, P3H-25-072 
  \end{flushright}
}

\title{
Power corrections to the production of a prompt photon in association with 
a jet in the $N$-jettiness 
slicing scheme at NLO QCD
}
\author[a]{Prem Agarwal,}
\author[a]{Kirill Melnikov,}
\author[a,b]{Ivan Pedron}

\affiliation[a]{\TTPaff}
\affiliation[b]{\IAPaff}

\abstract{ We compute  the 
next-to-leading-power corrections in the $N$-jettiness variable to  the production of a prompt photon and a jet at next-to-leading order in perturbative QCD in the $q \bar q$ annihilation channel. We employ the $k_\perp$ 
jet algorithm and assume that the $N$-jettiness value divided by the jet transverse momentum is the smallest parameter in the problem; in particular it should be small compared to the jet radius $R$. 
}

\begin{document}

\maketitle 

\section{Introduction}

A robust  description of hard scattering processes at the LHC requires 
significant  theoretical innovations. 
Such innovations encompass many  different aspects of collider 
theory including  computation of scattering amplitudes at high orders of QCD and electroweak perturbation theory~\cite{
Gehrmann:2015bfy,Chicherin:2018mue,Abreu:2018rcw,Abreu:2021oya,Chicherin:2018old,Abreu:2020jxa,Canko:2020ylt,Abreu:2021smk,Kardos:2022tpo,Abreu:2023rco,Chicherin:2021dyp,DeLaurentis:2023izi,DeLaurentis:2023nss,Agarwal:2024jyq,Badger:2024dxo,Agarwal:2023suw,FebresCordero:2023pww,DeLaurentis:2025dxw,Gehrmann:2018yef, Bargiela:2021wuy,Caola:2021rqz,Caola:2021izf,Canko:2021xmn,
Gehrmann:2023jyv, Chen:2025utl, Gehrmann:2024tds,Liu:2024ont,Henn:2023vbd,DiVita:2014pza,Long:2024bmi},
development of subtraction and slicing schemes for real-radiation contributions \cite{ Gehrmann-DeRidder:2005btv, Caola:2017dug, Currie:2013vh, DelDuca:2016csb, DelDuca:2016ily, Czakon:2010td, Czakon:2011ve, Czakon:2014oma, 
Catani:2007vq,
Jouttenus:2011wh, Gaunt:2015pea, 
Cacciari:2015jma,
Magnea:2018hab,  Bertolotti:2022aih, Stewart:2010tn, Buonocore:2023rdw,Fu:2024fgj,Boughezal:2015dva,Devoto:2023rpv,Devoto:2025kin,Devoto:2025jql,Fox:2024bfp,Gehrmann:2023dxm},  refinement of  parton shower 
programs~\cite{
vanBeekveld:2022zhl,vanBeekveld:2025lpz,FerrarioRavasio:2023kyg,Forshaw:2025bmo,Forshaw:2025fif,Altmann:2025yip,El-Menoufi:2024sys,Herren:2022jej} and their interfaces to fixed-order calculations \cite{Skands:2023lkp,Monni:2019whf,Alioli:2021qbf,Alioli:2025hpa},   as well as  advances in understanding  non-perturbative hadronization effects \cite{Hoang:2024nqi, Gieseke:2025mcy}. 

\emph{Power corrections} in the  slicing schemes are  one 
aspect  of the theoretical description of hard scattering processes  at the LHC where  further progress  is  desirable. Such  corrections  appear because of the very nature of slicing computations, where one separates  the phase space for a process with  $N$
final-state particles or jets, into   
phase-space regions where \emph{all} $N$
partons are  resolved, and 
regions where only a smaller number of partons or jets are resolved.  

This  separation  requires a  resolution variable. If the resolution variable is taken to be very small, dependencies of 
cross sections on the resolution variable, 
originating from resolved and unresolved regions,
  follow the  double-logarithmic pattern,  which is typical to  radiative corrections in QCD.  Such  dependencies 
on resolution variables are nearly universal and  are very well understood.  However, for a better matching between the resolved and unresolved contributions to cross sections, it is beneficial to expose  the dependence 
of  the unresolved 
contribution on the resolution variable \emph{beyond} the leading terms. Such subleading terms are referred to as power corrections. 

Power corrections for different resolution variables were investigated in a large number of publications in recent years ~\cite{Cieri:2019tfv,Boughezal:2016zws, DelDuca:2017twk, Boughezal:2018mvf, Moult:2016fqy, Moult:2017jsg, Ebert:2018lzn, Boughezal:2019ggi, vanBeekveld:2019prq,Oleari:2020wvt,Vita:2024ypr,Pal:2023vec,Pal:2024eyr,Beneke:2018gvs,Beneke:2019oqx,Bonocore:2015esa,Bonocore:2016awd,Broggio:2021fnr,Broggio:2023pbu,Ebert:2018gsn,Laenen:2020nrt,Pal:2025ffp,Czakon:2023tld}.
The majority of these papers  focused on contributions that are either logarithmically enhanced for small values of a resolution variable, or originate from 
emissions of soft gluon only.  It is therefore  unclear how  to extend these studies to  arbitrary collider processes and, in particular, to go beyond the soft limit in a general way.  Amusingly, this problem exists even at the next-to-leading order in perturbative QCD,  where one might  have thought that  everything is well-understood by now.

In Ref.~\cite{Agarwal:2025dvo} we  presented  a methodology to compute  subleading power corrections to {\emph arbitrary colorless} final states at NLO QCD using the $N$-jettiness resolution variable \cite{Stewart:2010tn}. The next natural step is to 
remove the restriction on the final states, and design a way to compute power corrections in 
the $N$-jettiness variable for arbitrary processes at colliders.  To simplify this step,  in this paper we consider such corrections to the process where  a prompt photon is produced in association with a jet. 
Furthermore, since very little is known about power corrections to processes with jets,\footnote{We are aware of a single paper 
\cite{Boughezal:2019ggi}
where  power corrections to the production of a vector boson in association with a jet are studied.  However, in that  paper only logarithmically-enhanced contributions to power corrections have been  computed, and an unconventional jet algorithm  has been  employed,  to  simplify the computation.}
we decided to first consider a single partonic channel 
$q \bar q \to \gamma + g$. For this channel,  the so-called photon isolation \cite{Frixione:1998jh} is not needed, and we can focus on the central question that we want to discuss, namely how the presence of a  jet algorithm affects the computation of power corrections. The process 
$q \bar q \to \gamma + g$
is well suited to study this question,  since it is sufficiently   simple, and we can directly work with the relevant matrix elements  to understand power corrections to  the partonic cross section. 

We  note that power-suppressed corrections arise also from observables or the selection cuts that define fiducial cross sections.  In this paper, we assume that observables are such that their dependence on the $N$-jettiness variable is analytic.\footnote{As we demonstrate below, this class of observables includes the  fully-realistic  inclusive sequential $k_\perp$ jet algorithm \cite{Ellis:1993tq,Catani:1993hr}.} 
It is known, however, that this is not always the case and that observables exist which induce a non-analytic  dependences of fiducial cross sections on the resolution variable
~\cite{Ebert:2019zkb,Salam:2021tbm,Ebert:2020dfc,Vita:2024ypr,Campbell:2024hjq, Alioli:2025hpa} which enhances the  power-suppressed contributions.  

The rest of the paper is organized as follows. In Section~\ref{sect2} we explain  how the presence of a jet in the final state affects the computation of  power corrections, and define  quantities that we use in the remainder of the paper.   We also summarize the  method for calculating  power corrections developed in Ref.~\cite{Agarwal:2025dvo}, that we employ in this paper. 
In Section~\ref{sect3}
we compute the power corrections in the $N$-jettiness
variable 
to the process  $q \bar q \to  \gamma+{\rm jet}$. We  investigate various soft and collinear contributions, as well as  subtleties  related to differences between cases when partons are clustered into a jet and cases when they are not. 
We also discuss the validation of our results in Section~\ref{sect3}.
We conclude in Section~\ref{sect:conclusions}.  Some technical details are discussed in  appendices.  

\section{General remarks}
\label{sect2}

We study the  production of a prompt photon and a jet in hadron collisions, $pp \to \gamma+j$. We  focus 
on the $q \bar q \to \gamma + g$ partonic channel, and do not consider any other partonic channels in this paper. We imagine that  the $N$-jettiness slicing scheme \cite{Stewart:2010tn,Boughezal:2015dva,Gaunt:2015pea}
is employed for computing the NLO QCD corrections.  An important ingredient for calculations  in this  scheme is 
 the   differential cross section for final states with   ${\cal T}_1 < \tau_{\rm cut}$, where 
${\cal T}_1$ is the $N$-jettiness variable, and 
$\tau_{\rm cut}$ is a small quantity.   To allow the 
choice of  somewhat larger values of $\tau_{\rm cut}$ in  practical computations, we need to construct an  expansion of the 
$q \bar q \to \gamma + j$ cross section through  first \emph{subleading} power in the one-jettiness  variable ${\cal T}_1 \sim \tau_{\rm cut}$.

To define a final-state jet, we require  a  jet algorithm.  We will consider the so-called inclusive  sequential 
$k_\perp$ jet 
algorithm \cite{Ellis:1993tq, Catani:1993hr}.\footnote{We note that our results can be used, without any modification,  for 
the anti-$k_\perp$ jet clustering algorithm
as well. 
} We will
describe how it works before addressing the  question of how it impacts the calculation of power corrections. 

To this end, we introduce  two phase-space ``distances'' 
\be
d_{ij} = {\rm min}(k_{\perp,i}^2,k_{\perp,j}^2) \;\frac{R^2_{ij}}{R^2},\;\;\; d_{iB} =  k_{\perp,i}^2,
\ee
where  $k_{\perp,x}$ is  the transverse momentum of a parton $x \in (i,j)$  defined with respect  to the collision axis.  We note that $d_{ij}$ and $d_{iB}$ measure  distances between the final-state partons $i$ and $j$, and between 
the final-state parton $i$ and the  beam axis, respectively. 
For the quantity $R_{ij}$, one typically takes 
\be
R_{ij}^2 = (\eta_i - \eta_j)^2 + f_\varphi^2(\varphi_i - \varphi_j), 
\label{eq1.1}
\ee
where $\eta_{i,j}$ are pseudo-rapidities of the two partons $i$ and $j$,  and $f_{\varphi}$ is a function of their azimuthal angles $\varphi_{i,j}$.  We  choose 
\be
f_\varphi(\varphi_i-\varphi_j)
= \arccos(\cos(\varphi_i-\varphi_j) ),
\ee
since this maps  the difference of two azimuthal angles onto the $[0,\pi]$ interval, independent of how azimuthal angles are parametrized.

To apply the  jet algorithm to a  set  of final-state partons  $P_N = \{1,2,3,..,N\}$, we start by computing two lists. One  of them  is  composed of $d_{ij}$'s calculated   for each
$(ij)$ pair from $P_N$, and the second -- of 
$d_{iB}$ for each $i \in P_N$.  We then compare  the minimal values of the two lists
\be
d_{\rm min} = {\rm min} \left [ {\rm min} \{ d_{ij} \}, {\rm min} \{d_{iB} \} \right ]. 
\ee
If $d_{\rm min}$ is the minimum of the  $\{d_{ij}\}$ list, the two partons $i$ and $j$ are removed from $P_N$ and replaced there by a new parton $ij$ whose
momentum is $p_{ij} = p_i + p_j$. If, however, the minimum is provided by the $\{d_{iB}\}$ list,  the parton $i$ is removed from
the list $P_N$   and added to the   list of \emph{jets} $P_J$ that is empty
at the start of this procedure.    We  continue  this process  until   no  partons 
are left in the list $P_N$.
Finally, all jets 
with the transverse momentum lower 
than a pre-selected  value $p_{\perp, \rm cut}$ are removed from the list of jets. 
Once this is done, we associate a  definite  number of jets with the partonic final state described by the original list $P_N$.

We continue with the discussion of  what this algorithm 
implies for the  computation
 of the power corrections.  
At leading order, we apply it to the  partonic process 
\be
q_a + \bar q_b \to \gamma  + g_\xa.
\label{eq2.5}
\ee
Therefore, the list of partons  consists of a single gluon $g_\xa$. This parton is moved to the list of jets immediately,   and if its transverse momentum exceeds $p_{\perp, \rm cut}$, it is identified with  a jet.
Once the jet is identified, we compute the one-jettiness and find 
\be
{\cal T}_1 = {\rm min} \left \{ \frac{2 p_{\xa} p_a }{P_a}, \frac{2 p_{\xa} p_b }{P_b}, \frac{2 p_\xa p_J}{P_J} \right \}  =0, 
\ee
because in this case $p_J = p_\xa$. 
\\

At next-to-leading order, we have to consider the  process\footnote{We  do not need to consider  the virtual corrections to the process in Eq.~(\ref{eq2.5})
because they will only contribute at ${\cal T}_1=0$.
} 
\be
q_a + \bar q_b \to \gamma + g_\xa + g_\yb,
\ee
and apply the jet algorithm to two gluons 
$g_{\xa}$ and $g_{\yb}$. To simplify further steps, it is convenient to order 
the two gluons  in the transverse momenta,  
and label them in such a way that $p_{\perp,\xa} >  p_{\perp,\yb}$.   The starting point for the application of the jet algorithm is the list 
$P_2 = \{\xa, \yb\}$. We have to find the minimum of $\{d_{\xa \yb}, d_{\xa B}, d_{\yb B}\}$. Thanks to the transverse momentum ordering, this minimum
is $d_{\xa \yb}$ if $R_{\xa \yb} < R$,  and $d_{\yb B}$ if $R_{\xa \yb} > R$.  Then, in the first (clustered) case, we have a one-jet event with the jet momentum
$p_J = p_\xa + p_\yb$ provided that $p_{\perp,J} > p_{\perp,\rm cut}$,  and in the second (unclustered) case we have, potentially, two jets with the transverse momenta
$p_{\perp,\yb}$ and $p_{\perp,\xa}$. To have a one-jet event we require $p_{\perp,\yb} < p_{\perp,\rm cut}$
and $p_{\perp,\xa} > p_{\perp, \rm cut}$.

Following  this discussion, we can make  the NLO QCD real-emission contribution to the $\gamma+j$ production  explicit.  To simplify the notation,  we write the one-jettiness
variable with an argument which refers to the momentum of the jet used in its definition, i.e. 
\be
   {\cal T}_1(p_J) = \sum \limits_{i \in \{ \xa, \yb \} }
   {\rm min} \left \{ \frac{2 p_i p_a }{P_a}, \frac{2 p_i  p_b }{P_b}, \frac{2 p_i p_J}{P_J} 
   \right \}.
\ee
As we just described, the jet momentum depends on whether gluons are clustered into a jet or not.  We find 
\be
\begin{split} 
  \frac{{\rm d} \sigma_{\cal O}}{{\rm d} \tau}  & = {\cal N}^{-1} \int  {\rm d} \Phi(p_a,p_b|p_\xa,p_\yb,p_\gamma) \; |{\cal M}|^2(p_a,p_b;p_\xa,p_\yb,p_\gamma)
   \; \theta(p_{\perp,\xa} - p_{\perp,\yb}) \\
   & \times \Big \{   \theta(R_{\xa \yb} - R) \; \theta(p_{\perp,\xa} - p_{\perp,\rm cut}) \; \theta(p_{\perp,\rm cut} - p_{\perp,\yb}) \;
   \delta( \tau - {\cal T}_1(p_\xa) ) \; {\cal O}(p_\xa,p_\gamma)
   \\
   & + \theta(R - R_{\xa \yb} ) \; \theta(p_{\perp,[\xa \yb]} - p_{\perp,\rm cut}) \;
   \delta( \tau - {\cal T}_1(p_{[\xa \yb]})) \; {\cal O}(p_{[\xa \yb]},p_\gamma)
   \Big \},
\label{eq1.6}
\end{split} 
\ee
where  ${\cal N}$ is the normalization factor,  
${\cal O}$ is an observable that depends on the jet  momentum and the momentum of the photon, and ${\rm d}\Phi$ is the phase space that will be  defined in the next section. We use $p_{[\xa \yb]}$ 
to denote the sum of the gluon momenta,   $p_{[\xa \yb]}= p_\xa + p_\yb$.

To simplify the notation, we will absorb 
the $\theta$-functions, that ensure that the transverse momentum of the jet is larger than the transverse-momentum cut, into the definition of the observable ${\cal O}$.
 Hence,  from now on, we will only write explicitly  the $R$-dependent $\theta$-functions  from the jet algorithm, 
 as well as the 
 $\theta$-function that ensures that the transverse momentum of the parton $\yb$
 is small.

There is an important difference in the one-jettiness functions that appear in the two terms in the integrand in Eq.~(\ref{eq1.6}).
In the first term, $p_J = p_\xa$ and therefore
\be
   {\cal T}_1(p_\xa) =
         {\rm min} \left \{ \frac{2 p_\yb p_a }{P_a}, \frac{2 p_\yb  p_b }{P_b}, \frac{2 p_\yb p_\xa}{P_J} \right \}.
         \label{eq2.10}
\ee
In the second term, $p_J = p_{[\xa \yb]}$, and we  find a more complicated expression for the one-jettiness function
\be
\begin{split}
   {\cal T}_1(p_{[{\xa \yb}]}) & = \sum \limits_{i \in \{ \xa, \yb \} }
   {\rm min} \left \{ \frac{2 p_i p_a }{P_a}, \frac{2 p_i  p_b }{P_b}, \frac{2 p_i p_{[\xa \yb]}}{P_J} \right \} 
 \\
  &  =    {\rm min} \left \{ \frac{2 p_\yb  p_a }{P_a}, \frac{2 p_\yb   p_b }{P_b}, \frac{2 p_\yb p_\xa}{P_J} \right \}
      + {\rm min} \left \{ \frac{2 p_\xa p_a }{P_a}, \frac{2 p_\xa   p_b }{P_b}, \frac{2 p_\xa p_\yb  }{P_J} \right \}.
      \label{eq2.11}
\end{split} 
   \ee
To compute power-suppressed one-jettiness corrections, we need to analyze different contributions to Eq.~(\ref{eq1.6}), finding 
kinematic regions where the one-jettiness function defined in Eqs~(\ref{eq2.10},\ref{eq2.11}) is small.

If the two 
partons $\xa$ and $\yb$ have  generic momenta, ${\cal T}_1$ cannot be small. For this to occur, partons 
$\xa$ and $\yb$ should have special, singular  kinematics. 
In general,  given  that $p_{\perp, \xa} > p_{\perp, \yb}$ and at least one jet is required, 
there are four options for the parton $\yb$:
      \begin{itemize}
      \item  $\yb$ is collinear to $a$, $b$ or $\xa$;
        \item $\yb$ is soft.
        \end{itemize}

We will continue with the discussion of these  cases separately. 
We choose the reference frame where momenta of partons $a$ 
and $b$ are  along the $z$ axis, and we denote 
a polar angle of a parton $x$ by  $\theta_x$.

    \paragraph{ The collinear case   $\yb || a$}  In this case, the energy of the parton $\yb$ is large, $E_\yb \sim E_a \sim E_b$,  but the  polar angle is small, $\theta_{\yb} \sim \sqrt{\tau P_J/E_a^2}$.  At the same time the four-momentum of the parton $\xa$ is generic, i.e. $E_{\xa} \sim E_a$
    and $\theta_{\xa} \sim 1$. 
     Clustering of  $\xa$ and  $\yb$ into a jet is impossible because the rapidity of the  parton $\yb$
    is very large
    \be
  \eta_\yb = \frac{1}{2} \ln \frac{1+\cos \theta_{\yb}}{1 - \cos \theta_{\yb}} \sim -\frac{1}{2} \ln \frac{\tau P_J}{E_a^2} .
    \ee
    Therefore,  
    \be
R_{\xa \yb}^2 \sim \ln^2  \left(\frac{ \tau P_J}{E_a^2} \right),
    \ee
    and,
    as long as 
    \be
 \ln \left ( \frac{E_a^2}{\tau P_J}  \right ) \gg R,
 \label{eq2.14}
\ee
clustering of partons $\xa$ and $\yb$ into a single jet 
does not   occur.\footnote{ Another condition on the jet radius that restricts it from \emph{above}, is derived later.}  Hence, in this case we can write 
\be
\begin{split} 
  \frac{{\rm d} \sigma^{ca}_{\cal O}}{{\rm d} \tau} 
  = & {\cal N}^{-1} \int  {\rm d} \Phi(p_a,p_b|p_\xa,p_\yb,p_\gamma) |{\cal M}|^2(p_a,p_b;p_\xa,p_\yb, p_\gamma) \\
  & \times
\delta \left ( \tau - \frac{2 p_a p_\yb}{P_a} \right ) {\cal O}(p_\xa,p_\gamma).
\end{split} 
\label{eq2.15a}
\ee
Note that we have dropped the constraint 
on the transverse momentum of the parton $\yb$ because $p_{\perp, \yb} \sim \sqrt{\tau P_a }
$ and, as long as $\tau$ is small  and $p_{\perp,\rm cut} \sim {\cal O}(E_a)$, the transverse momentum of the parton $\yb$ cannot exceed the cut value.

\paragraph{The collinear case $\yb || b$} This case is analogous to the $\yb || a$  case.  Hence, without further discussion, we write 
\be
\begin{split} 
  \frac{{\rm d} \sigma^{cb}_{\cal O}}{{\rm d} \tau}  &
  = {\cal N}^{-1} \int  {\rm d} \Phi(p_a,p_b|p_\xa,p_\yb,p_\gamma) \;  |{\cal M}|^2(p_a,p_b;p_\xa,p_\yb, p_\gamma)
\\
 &  \times 
\delta \left ( \tau - \frac{2 p_b p_\yb}{P_b} \right ) {\cal O}(p_\xa,p_\gamma).
\end{split} 
\label{eq2.17}
\ee

\paragraph{The collinear case $\xa || \yb$} This case corresponds to the final-state collinear configuration.  Computing the invariant mass of partons $\xa$ and $\yb$  in the collinear approximation, we obtain  
\be
s_{\xa \yb} \approx  E_\xa E_\yb \left ( \left ( \theta_\xa - \theta_\yb \right )^2 + \sin^2 \theta_{\xa}  
\left ( \varphi_{\xa} - \varphi_{\yb} \right )^2 \right ) 
\approx  p_{\perp,\xa} p_{\perp,\yb} R_{\xa \yb}^2.
\ee
Using the jettiness constraint, we estimate that in  the collinear $\xa || \yb$ case, the $s_{\xa \yb}$ invariant mass becomes 
\be
s_{\xa \yb} \sim \tau P_J.
\ee
Hence, 
\be
R_{\xa \yb}^2  
\sim \frac{\tau P_J}{p_{\perp, \xa} p_{\perp, \yb}} 
\sim  \frac{\tau P_J}{p_{\perp, \rm cut}^2}
\ll R^2.
\label{eq2.19}
\ee
and the two partons are clustered into a single jet.\footnote{We note that, 
under the assumption that 
$p_{\perp,\rm cut} \sim E_a \sim P_J$, 
Eqs~(\ref{eq2.14}) and (\ref{eq2.19}) \emph{together} imply that the jet radius should satisfy the following constraint  $\sqrt{\tau/p_\perp} \ll R 
\ll \ln p_\perp/\tau$. } 
The expression for the cross section reads 
\be
\begin{split} 
  \frac{{\rm d} \sigma_{\cal O}^{\xa \yb}}{{\rm d} \tau}  & =  
  \frac{{\cal N}^{-1}}{2} 
  \int  {\rm d} \Phi(p_a,p_b|p_\xa,p_\yb,p_\gamma) \;  |{\cal M}|^2(p_a,p_b;p_\xa,p_\yb, p_\gamma)
    \\
&  \times 
   \delta \left ( \tau - \frac{4 p_\xa p_\yb}{P_J} \right ) {\cal O}(p_{[\xa \yb]},p_\gamma),
\end{split} 
\ee
We note that we have introduced the factor $1/2$, and used  the   $\xa \leftrightarrow \yb$ replacement symmetry in this kinematic configuration, to remove the  $p_{\xa, \perp} > p_{\yb, \perp}$ condition from the integrand. 

\paragraph{The soft case $E_\yb \to 0$ }  Finally, we need 
to discuss the soft case where  $E_{\yb} \sim \tau$. Then  
$p_{\perp,\yb} \ll p_{\perp,\xa}
\sim p_{\perp,{\rm cut}},  $ 
and conditions  that ensure  these requirements can be dropped. At the same time, since the soft gluon $\yb$ can be emitted at an arbitrary angle, it is impossible to say a priori whether it will be clustered into a jet together with $\xa$,  or not. 
Because of this, we write 
 the soft contribution in such a way, that both clustered and non-clustered cases can be described. The soft contribution reads 
\be
\begin{split} 
  \frac{{\rm d} \sigma^{s}_{\cal O}}{{\rm d} \tau}  & =
       {\cal N}^{-1}
       \int {\rm d} \Phi(p_a,p_b|p_
       \xa,p_\yb,  p_\gamma) |{\cal M}|^2(p_a,p_b;p_\xa,p_\yb,p_\gamma)
  \\
   & \times  \Theta(R_{\xa \yb}, q_j)  \;
  \delta( \tau - {\cal T}_1(q_j) ) \; {\cal O}(q_j,  p_\gamma),
\end{split}
\label{eq2.21}
\ee
where $q_j$ is the momentum 
of the identified jet, and $\Theta(R_{\xa \yb},q_j)$ is the remnant of the angular distance of the jet algorithms  defined as follows
\begin{equation}
\begin{split}
\Theta(R_{\xa \yb},q_j)  = \theta(R_{\xa \yb} - R)\; \delta_{q_j,p_\xa}
+\theta(R-R_{\xa \yb}) 
\; \delta_{q_j,p_{[\xa \yb]}}.
\end{split}
\end{equation}
The  one-jettiness function reads 
\be
   {\cal T}_1(q_j) = {\rm min}
   \left \{ \frac{2 p_a p_\yb}{P_a},\frac{2 p_b p_\yb}{P_b},\frac{ 2 p_\xa p_\yb }{P_J} \right \} 
   + \delta_{q_j, p_{[\xa \yb]} } \; \frac{2 p_\xa p_\yb}{P_J}.
   \label{eq2.23}
   \ee
The last term distinguishes between the clustered and the non-clustered cases. 

\paragraph{Computational  strategy}

To compute the various contributions, we adopt the strategy discussed in 
Ref.~\cite{Agarwal:2025dvo}, where we constructed Lorentz transformations for different cases,  used them to  factorize the phase space 
for the photon  and the two partons with the power accuracy, and expanded the squared matrix element   and the observable functions around soft and collinear limits. In Ref.~\cite{Agarwal:2025dvo} we developed a process-independent procedure to expand the matrix element 
for the production of a color-singlet final state.  For simplicity, in this paper  we make use of the  explicit form of the matrix element for the  $q \bar q \to \gamma+g + g$ process,  to construct an expansion in the  soft and  collinear limits. 

The required Lorentz transformations  for the cases 
$\yb ||a$ and $\yb || b$, as well as for the case when the  parton  $\yb$ is soft, are discussed in detail in Ref.~\cite{Agarwal:2025dvo}. The new technical element required here is  the momenta  mappings   for  the collinear $\xa || \yb$ case.  We describe these mappings  in Appendix~\ref{sec:finalboost}.

\section{Power corrections  to the $\gamma$+jet production in the $q \bar q \to g \gamma$ channel}
\label{sect3}

\subsection{Leading order}

We consider the partonic process 
\be
q(p_a) + \bar q(p_b) \to 
\gamma(p_\gamma) + g(p_j),
\label{eq4.1}
\ee
and associate the final-state gluon with a jet. 
The differential cross section of the 
process in Eq.~(\ref{eq4.1}) 
reads 
\be
   {\rm d} \sigma_0 =
   \frac{ N_c Q_q^2 (e g_s)^2 }{ 2 s N_c^2 } \int {\rm d} \Phi^{ab}_{\gamma j} 
   \;
  \sum \limits_{\rm col,pol}^{}
   \frac{|{\cal M}_0|^2(p_a,p_b,p_j,p_\gamma)}{4 N_c (Q_q e g_s)^2} \;    {\cal O}(p_j,p_\gamma),
\ee
where $Q_q$ is the quark electric charge in units of the positron charge $e$, 
$g_s$ is the (bare) strong coupling constant, $N_c=3$ is the number of colors,   
$s = 2 p_a \cdot p_b$ and 
${\cal O}(p_j,p_\gamma)$ is  the infrared-safe observable 
that depends on the momenta of the jet and the photon. Furthermore,  
\be
{\rm d} \Phi^{ab}_{\gamma j} 
= [d p_j][d p_\gamma](2 \pi)^d \delta^{(d)}(p_a + p_b -p_\gamma - p_j),  
\ee
is the phase space\footnote{Throughout the paper, we employ dimensional regularization and work in $d=4-2\ep$ dimensions.}  with 
\be
[{\rm d} p_x] = \frac{{\rm d}^d p_x}{(2\pi)^{d-1}} \; \delta_+(p_x^2).
\ee

We employ the  Sudakov decomposition of the photon 
and jet momenta to 
parametrize the Born phase space. We use  momenta of the  incoming partons to define the basis vectors.   Then, 
\be
\begin{split} 
& p_j = \beta p_a + (1-\beta) p_b 
- \sqrt{s \beta (1-\beta)} \; n_\perp,
\\
& p_\gamma = (1-\beta) p_a + \beta  p_b 
+ \sqrt{s \beta (1-\beta)} \; n_\perp,
\end{split} 
\label{eq4.4}
\ee
with $\beta \in [0,1]$, $p_{a,b} \cdot n_\perp = 0$ and $n_\perp^2 = -1$.
The transverse momentum of the jet reads 
\be
|p_{\perp,j}| = \sqrt{s \beta (1-\beta)}.
\ee
We use  Eq.~(\ref{eq4.4}), to write the Born phase space as follows
\be
{\rm d} \Phi_{\gamma j}^{ab} = 
\frac{s^{-\ep}
\Omega^{(d-2)}}{4 (2\pi)^{d-2}} \; {\rm d} \beta \;  [{\rm d} \Omega^{(d-2)}] \;
\beta^{-\ep} (1-\beta)^{-\ep}
=
\frac{1}{8 \pi } {\rm d} \beta \; \frac{{\rm d} \varphi}{2 \pi} +{\cal O}(\ep),
\label{eq3.7}
\ee
where $\Omega^{(d-2)}$ 
is the solid angle in  $d-2$ dimensions and 
$[{\rm d} \Omega^{(d-2)}] = {\rm d} \Omega^{(d-2)}/\Omega^{(d-2)}$. 
The azimuthal angles  in  ${\rm d} \Omega^{(d-2)}$  
  parametrize the direction of the vector $n_\perp$ in the ($d-2$)-dimensional space orthogonal to $p_{a,b}$.

The appropriately normalized squared matrix element  for the process in 
Eq.~(\ref{eq4.1}),  summed over polarizations and colors reads 
\be
\sum \limits_{\rm pol,col} \frac{|{\cal M}_0|^2(p_b,p_a;p_\xa,p_\gamma)}{4 N_c 
(Q_q e g_s)^2 }  
= C_F \left[ \frac{(1-\ep)}{2} \left( \frac{t}{u} +  \frac{u}{t} \right) - \ep \right]
 = C_F \frac{( 1 - 2 \beta + 2 \beta^2 - \ep)}{2\beta (1-\beta)},
\label{eq4.7}
\ee
where $t = -2 p_a \cdot p_\gamma$, $u = -2 p_b \cdot p_\gamma$.  
Finally, using 
the above  ingredients, 
we write the leading order differential cross section as  
\be
{\rm d} 
\sigma_0 =  \bar \sigma_0 \;
{\rm d} \Phi_{\gamma j}^{ab} \; \frac{( 1 - 2 \beta + 2 \beta^2 - \ep)}{2\beta (1-\beta)}, 
\ee
where
\be
\bar \sigma_0 = \frac{16 \pi^3  C_F Q_q^2 \; \alpha_{\rm QED} \;[\alpha_s] }{s N_c},
\ee 
with 
\be
[\alpha_s] = \frac{g_s^2 \Omega^{(d-2)}}{2(2\pi)^{d-1}} = \frac{\alpha_s}{2 \pi} + {\cal O}(\ep).
\ee

Having discussed the cross section for the Born process, we proceed with the computation of the power corrections in 
the one-jettiness variable.
As pointed out  in Section~\ref{sect2}, several contributions need to be considered.  
We will start with the discussion of the soft case, 
and continue with  the collinear ones.

 \subsection{The  soft contribution}
\label{sec3.2}

We consider   the case when the parton $\yb$ is soft, which means that its energy is of order $\tau$. We note that this kinematic configuration has to be considered for the case   when partons $\xa$ and $ \yb$ are clustered into a jet,  and for the case when they are not. 
Our starting point is Eq.~(\ref{eq2.21}) 
that we repeat here for convenience
\be
\begin{split} 
  \frac{{\rm d} \sigma^{s}_{\cal O}}{{\rm d} \tau}  & =
       {\cal N}^{-1}
       \int {\rm d} \Phi(p_a,p_b|p_
       \xa,p_\yb, \tilde p_\gamma) |{\cal M}|^2(p_a,p_b;p_\xa,p_\yb, \tilde p_\gamma)
  \\
   & \times  \Theta( R_{\xa \yb}, q_j)  \;
  \delta( \tau - {\cal T}_1(q_j) ) \; {\cal O}(q_j, \tilde p_\gamma).
\end{split}
\label{eq4.38}
\ee

We note that the  normalization factor 
in Eq.~(\ref{eq4.38}) coincides with the one for  the  leading order  process  $q \bar q \to g + \gamma$. 
This means that 
\be
{\cal N}^{-1} \; {\rm d} \Phi^{ab}_{\gamma j} \;
|{\cal M}_0(p_a,p_b,p_j,p_\gamma)|^2 = 
\bar \sigma_0 \; {\rm d} \Phi^{ab}_{\gamma j} \; 
\frac{(1-2 \beta + 2 \beta^2-\ep)}{2 \beta (1-\beta) }, 
\label{eq4.41}
\ee
and we will use this equation 
for simplifying  some computations in what follows.  

We also note that 
we have written  the photon momentum in Eq.~(\ref{eq4.38}) as $\tilde p_\gamma$. This is done on purpose since, because of  Lorentz transformations, this momentum will be  redefined as we proceed with the calculation,   and we would like to reserve the notation $p_\gamma$ for the photon momentum appearing in the final equations.

      The soft contribution corresponds to the scaling $p_\yb  \sim \tau$. For the sake 
      of convenience, 
     in what follows 
      we will 
      refer to $p_\yb$ as $k$. 
  To construct the expansion around the soft limit, we define the four-momentum 
   \be
   P_{ab} = p_a + p_b, 
   \ee
and perform a boost and a rescaling 
to remove  the  momentum $k$ from the energy-momentum conservation constraint $p_a+p_b = p_\xa + k + p_\gamma$, which is  implicitly present   in Eq.~(\ref{eq4.38}). We write 

   \be
\lambda P_{ab}^\mu = 
   [\Lambda_s]^\mu_{\nu}    \; (P_{ab} -k)^\nu.
   \ee
  The matrix   $\Lambda_s$ in the above equation is the Lorentz boost. The rescaling parameter 
   $\lambda$, computed through 
   first order in $k \sim \tau$, reads
   \be
\lambda \approx 1 - \frac{P_{ab} \cdot k}{P_{ab}^2}.
   \ee

   Performing the boost, and using the   phase-space modification in the soft limit computed in Ref.~\cite{Agarwal:2025dvo}, we  find
\be
\begin{split} 
  \frac{{\rm d} \sigma^{s}_{\cal O}}{{\rm d} \tau}  & =
       {\cal N}^{-1} \int  {\rm d} \Phi^{ab}_{j \gamma}
       \; 
[{\rm d} k] \left ( 1 + 2\ep \frac{P_{ab} \cdot k}{P_{ab}^2} \right ) 
   \; \Theta(  R_{\xa \yb}, q_j) \\
 & \times |{\cal M}|^2(p_a,p_b; \lambda \Lambda^{-1}_s p_j,k,\lambda \Lambda^{-1}_s p_\gamma) \; \delta( \tau - {\cal T}_1(q_j) ) \; {\cal O}(q_j,  \lambda \Lambda^{-1}_s p_\gamma),
\end{split}
\label{eq3.16}
\ee
where $p_\xa = 
\lambda \Lambda_s^{-1} p_j$, $\tilde p_\gamma = \lambda \Lambda_s^{-1} p_\gamma$ and 
$p_j^2 = p_\gamma^2 = 0$.
 Furthermore, \begin{equation}
[{\rm d}  k] =  {\rm d} \omega_k \; \omega_k^{1-2\ep} \frac{\Omega^{d-2}}{2 (2 \pi)^{d-1}} [{\rm d} \Omega_k ], \;\;\;  \left [ \Lambda^{-1}_s \right ] _{\mu \nu}= 
g_{\mu \nu}  -\frac{k^\mu P_{ab}^\nu - P_{ab}^\mu k^\nu}{P_{ab}^2} +{\cal O}(k^2).
\end{equation}

Since the jet momentum 
$q_j$ depends 
on whether partons are clustered into a jet   or not, there are two  possible ways for  
$q_j$ to transform 
under the soft 
boost and rescaling. They are  
\be
\begin{split} 
& q_j = p_{\xa} + p_{\yb} = \lambda \Lambda_s^{-1} p_j + k,~~~{\rm if~clustered},
\\
& q_j = p_{\xa} = \lambda \Lambda_s^{-1} p_j,~~~{\rm if~ not~clustered}.
\end{split}
\label{eq3.18}
\ee

The one-jettiness function also differs for the two cases. However, since  
\be
p_\xa \cdot p_\yb = (\lambda \Lambda_s^{-1} p_j ) \cdot k =  p_j\cdot k
+{\cal O}(k^3),
\label{eq3.20}
\ee
we find that the following equation holds  
\be
   {\cal T}_1(q_j) = {\rm min}
   \left \{ \frac{2 p_a k}{P_a},\frac{2 p_b k}{P_b},\frac{ 2 p_j k }{P_J} \right \} 
   + \delta_{q_j, p_{[\xa \yb]} } \; \frac{2 p_j k}{P_J},
   \ee
and no ${\cal O}(k)$ corrections appear in the expression for the one-jettiness. 
\\

To compute ${\rm d} \sigma^{s}/{\rm d \tau}$ through first subleading correction in $\tau$, we need to expand all the relevant quantities  in the 
integrand in 
Eq.~(\ref{eq3.16}) to first subleading  order in the gluon energy $\omega_k$. 
This includes the expansion 
of the matrix element squared,  the observable and also the function 
$\Theta(R_{\xa \yb},q_j)$ which gets modified because   the angular distance between partons $j$ and $k$,  and $\xa$ and $k$ is not the same.  We will start with the discussion of the matrix element. 

The next-to-soft correction to the squared matrix element  can be obtained  from the extension of  the Burnett-Kroll-Low theorem \cite{Burnett:1967km,Low:1958sn} to QCD. For the process $q \bar q \to \gamma + j$, such a study was performed 
in Ref.~\cite{vanBeekveld:2019prq}, where it was shown that, with the required accuracy, 
the squared matrix element  for this process can be  written 
in the following way
\be
\begin{split}
    & g_s^{-2} |{\cal M}|^2(p_a,p_b,p_\xa ,k,\tilde p_\gamma)   \approx  
    \\ & 
    \left(C_F - \frac{C_A}{2}\right)\frac{2 p_a \cdot p_b}{p_a \cdot k \; p_b \cdot k}
|{\cal M}_0( p_a + \delta p_{a,b},
    p_b + \delta p_{b,a}, p_\xa, \tilde p_\gamma)
    |^2\\
    & +\frac{C_A}{2} \frac{2 p_a \cdot p_\xa }{p_a \cdot k \; p_\xa  \cdot  k}
|{\cal M}_0(p_a + \delta p_{a,j}, p_b,
    p_\xa  - \delta p_{\xa ,a}, \tilde p_\gamma)|^2
    \\
   & +\frac{C_A}{2} 
   \frac{2 p_b \cdot p_\xa }{p_b \cdot k \; p_\xa  \cdot k}
|{\cal M}_0|^2(p_a, p_b + \delta p_{b,\xa },
    p_\xa  - \delta p_{\xa ,b} ,
    \tilde p_\gamma)|^2.
\end{split}
\label{eq3.21}
\ee
The momenta shifts 
in Eq.~(\ref{eq3.21})
read 
\be
\delta p_{l,m} = - \frac{1}{2} \left( k+ \frac{p_m \cdot k}{p_l \cdot p_m} \ p_l - \frac{p_l \cdot k}{p_l \cdot p_m}  p_m \right).
\label{eq3.22}
\ee
They satisfy the following equations
\be
\delta p_{l,m} + \delta p_{m,l} 
= -k,\;\;\; (p_l \pm \delta p_{l,m})^2 =
\pm 2 p_l  \cdot \delta p_{l,m} + 
{\cal O}(k^2) = {\cal O}(k^2). 
\ee
These equations ensure that with the  next-to-soft accuracy,  all momenta that appear in 
the matrix element ${\cal M}_0$ in 
Eq.~(\ref{eq3.21}) are on-shell, and 
that the momentum conservation is satisfied,  provided that equation 
\be
p_a + p_b = p_\xa  + \tilde p_\gamma + k,
\ee
holds.

According to Eq.~(\ref{eq3.16}), 
we need to compute the matrix element 
 for boosted 
 and rescaled 
momenta.  We will make use of the fact that the mass dimension of the 
$|{\cal M}_0|^2$ is zero (see Eq.~(\ref{eq4.7})), and that it is boost-invariant. Then, the following equation holds
\be
|{\cal M}_0(p_a, p_b, \lambda \Lambda_s^{-1}p_j, \lambda \Lambda_s^{-1} p_\gamma)
|^2
= |{\cal M}_0(\lambda^{-1} \Lambda_s p_a, \lambda^{-1} \Lambda_s p_b,p_j, p_\gamma)
|^2.
\ee
It is easy to check, using explicit 
formula for the boost and the rescaling that 
\be
\lambda^{-1} \Lambda_s p_a = p_a 
- \delta p_{a,b}, 
\;\;\;
\lambda^{-1} \Lambda_s p_b = p_b 
- \delta p_{b,a},
\ee
where $\delta p_{a,b}$ and 
$\delta p_{b,a}$ 
are defined in  Eq.~(\ref{eq3.22}). Then, through next-to-soft terms, Eq.~(\ref{eq3.21}) becomes 
\be
\begin{split}
    & g_s^{-2} |{\cal M}|^2(p_a,p_b,\lambda \Lambda_s^{-1} p_j,k, \lambda \Lambda_s^{-1} p_\gamma)   \approx  
    \\ & 
    \left(C_F - \frac{C_A}{2}\right)\frac{2 p_a \cdot p_b}{p_a \cdot k \; p_b \cdot k}
|{\cal M}_0( p_a,
    p_b , p_j, p_\gamma)
    |^2\\
    & +\frac{C_A}{2} \frac{2 p_a \cdot \lambda \Lambda_s^{-1} p_j}{p_a \cdot k \; p_j \cdot k }
|{\cal M}_0(p_a -\delta p_{a,b} + \delta p_{a,j}, p_b-\delta p_{b,a},
    p_j - \delta p_{j,a}, p_\gamma)|^2
    \\
   & + \frac{C_A}{2} 
   \frac{2 p_b \cdot \lambda \Lambda_s^{-1} p_j}{p_b \cdot k \; p_j \cdot  k}
|{\cal M}_0|^2(p_a-\delta p_{a,b}, p_b -\delta p_{b,a} + \delta p_{b,j},
    p_j - \delta p_{j,b} ,
    p_\gamma)|^2,
\end{split}
\label{eq3.27}
\ee
where we have used the fact that 
$k \cdot (\lambda \Lambda_s^{-1} p_j) = k \cdot p_j$ with the required accuracy, c.f. 
Eq.~(\ref{eq3.20}).
We stress that momenta in Eq.~(\ref{eq3.27})   
satisfy the leading-order energy-momentum conservation equation 
\be
p_a + p_b = p_j + p_\gamma.
\ee
Furthermore, we  note a peculiar fact that the momenta transformations  removed the next-to-soft correction from 
the $(a,b)$ dipole, whereas such corrections do  
remain in the 
$(a,j)$ and $(b,j)$ dipoles. 

Eq.~(\ref{eq3.27}) provides a 
suitable starting point for computing the required expansion of the real-emission squared matrix element  in the soft limit through subleading power. We 
use explicit form of the leading-order matrix element squared Eq.~(\ref{eq4.7}) and find 
\be
\begin{split}
    & \frac{\omega_k^2}{g_s^2} \;  |{\cal M}|^2(p_a,p_b,\lambda \Lambda_s^{-1} p_j,k, \lambda \Lambda_s^{-1} p_\gamma)  
\approx 
  |{\cal M}_0|^2(p_a,p_b,p_j,p_\gamma) \left ( S_1(\vec n_k) 
    + \frac{\omega_k}{\sqrt{s}} \; S_2(\vec n_k)
       \right ),
       \end{split}
    \label{eq3.29}
       \ee
where 
\be
\begin{split}
& S_1(\vec n_k) = 
     \left(C_F - \frac{C_A}{2} \right)\frac{2 \rho_{ab}}{\rho_{ak} \rho_{bk}  }
    + \frac{C_A}{2} 
      \left ( \frac{2 \rho_{aj}}{\rho_{ak} \rho_{jk}} 
      + \frac{2 \rho_{bj}}{\rho_{bk} \rho_{jk}}
      \right ),
\\
&  S_2(\vec n_k) = 
C_A \left ( 
\frac{\rho_{ab}}{\rho_{ak} \rho_{bk}}
-\frac{\rho_{aj}}{\rho_{ak} \rho_{jk}}
-\frac{\rho_{bj}}{\rho_{bk} \rho_{jk}} 
- \frac{2}{\rho_{jk}}
\right ),
       \end{split}
       \label{eq3.31}
       \ee
       and $\rho_{xy} = 
       1- \vec n_x \cdot \vec n_y$.
We stress that in deriving 
Eq.~(\ref{eq3.29}) 
no $\ep$-dependent terms have been neglected.

We continue with the discussion of a generic observable ${\cal O}$.
We remind the reader that
${\cal O}$ contains the constraint on the jet  transverse momentum, according to our convention. 
 To find how the observable is affected by the momenta transformation,
 we note that  
according to Eq.~(\ref{eq3.21}), to compute the jet momentum  we always need to transform  the harder gluon $\xa$, and then either combine it with a softer gluon or not.  
Since 
\be
\lambda \Lambda^{-1}_s  p_j = \left ( 1- \frac{P_{ab} \cdot k}{P_{ab}^2} \right ) p_j 
- k \frac{P_{ab} \cdot p_j}{P^2_{ab}} + P_{ab} \frac{k \cdot p_j}{P_{ab}^2},
\ee
and $P_{ab} \cdot p_j/P_{ab}^2 = 1/2$,
we find the following result for the two cases 
\be
\begin{split} 
&{\bullet \; \rm clustered:}~~q_j = p_\xa + p_\yb = \lambda \Lambda_s^{-1} p_j + k 
= 
\left ( 1- \frac{P_{ab} \cdot k}{P_{ab}^2} \right ) p_j 
+ \frac{1}{2} k + P_{ab} \frac{k \cdot p_j}{P_{ab}^2},
\\
&{\bullet \ \rm not~clustered:}~~q_j = p_\xa = \lambda \Lambda_s^{-1} p_j
= \left ( 1- \frac{P_{ab} \cdot k}{P_{ab}^2} \right ) p_j 
- \frac{1}{2} k + P_{ab} \frac{k \cdot p_j}{P_{ab}^2}.
\end{split}
\ee

We can now expand the observable to the desired order in the soft approximation 
\be
\begin{split}
{\cal O}(q_j, \tilde p_\gamma)
& = 
{\cal O}(p_j, p_\gamma) 
+ \sum \limits_{x \in j,\gamma}
\left ( -\frac{P_{ab} \cdot k}{P_{ab}^2}  p_x^\mu
- \frac{1}{2} k^\mu+ \frac{k \cdot p_x}{P_{ab}^2} P_{ab}^\mu
\right )
\partial_{p_x,\mu}  {\cal O}(p_j, p_\gamma) 
\\
& + \theta(R-R_{jk}) \; k^\mu \partial_{p_j,\mu}  {\cal O}(p_j, p_\gamma) 
+{\cal O}(k^2).
\end{split}
\label{eq3.34}
\ee
We emphasize that when  gluons are clustered,  the square of the jet momentum $q_j^2 \ne 0$ whereas $p_j^2 = 0$. Hence, when computing the derivative
$\partial_{p_j}$ 
on the right-hand side of the above equation, one should write the definition of the observable \emph{without assuming} $p_j^2 = 0$, calculate  the derivative, and take the $p_j^2 \to 0$ limit after that. This remark concerns, in particular, the 
dependence of the observable
${\cal O}$ on the transverse momentum of  the jet.\footnote{The transverse momentum of the jet can be defined through the following equation 
$p_{j,\perp} = \sqrt{2 (p_a p_j) (p_b p_j)/(p_a p_b) -p_j^2}$.}

It is useful to  rewrite 
Eq.~(\ref{eq3.34}) separating 
the energy of the gluon $\omega_k$ from its direction. We will work in the center-of mass frame of the partonic collision, where  partons $a,b$ are back-to-back and have equal energies.  We find 
\be
\begin{split}
{\cal O}(q_j, \tilde p_\gamma)
& = 
{\cal O}(p_j, p_\gamma) + \frac{\omega_k}{\sqrt{s}} 
\Bigg[ 
\sum \limits_{x \in j,\gamma} \left ( -p_x^\mu
-\frac{\sqrt{s}}{2} \hat{k}^\mu + \frac{\rho_{kx}}{2} \; P_{ab}^\mu
\right ) \partial_{p_x,\mu} 
\\
& + \theta(R-R_{jk}) 
\; \sqrt{s} \; \hat k^\mu \partial_{p_j,\mu}
 \Bigg] {\cal O}(p_j, p_\gamma),
\end{split}
\ee
where $\hat k^\mu = (1,\vec n_k)$.

\paragraph{Modification of the angular distance in the jet algorithm}

Similarly to the matrix element and the observable, 
in Eq.~(\ref{eq3.16}) we need to expand the $\Theta(R_{\xa \yb},q_j)$-function that determines whether the gluons are clustered into a jet or not. The power correction in this case is actually finite; for this reason it is useful to consider it separately.

The original $\eta-\varphi$ distance refers to   partons $\xa$ and $\yb$. We have identified $\yb$ with $k$, 
but the momentum of $\xa$ is expressed through the (large) momentum $p_j$ and additional $k$-dependent terms. Hence, as explained in Appendix~\ref{app:shift},
in the center-of-mass frame of the colliding partons $p_{a,b}$, the following relation holds 
\be
R_{\xa \yb} = R_{jk} + 
\frac{\omega_k}{\sqrt{s}} \;  {\cal R}_{jka},
\ee
where $\omega_k \sim \tau$ is the (small) energy of the gluon $k$.
Indices of the  function ${\cal R}$ indicate that it depends on $\vec n_j, \vec n_k$ 
and $\vec n_a$.
Explicitly, this function reads
\be
{\cal R}_{jka} = \frac{1}{\sin^2 \theta_j}
[\vec n_k \times \vec n_j]
\cdot 
\left ( 
\frac{\partial R_{jk} }{\partial \varphi_j} \;  \vec n_a 
- \frac{\partial R_{jk}}{\partial \eta_j} \;  [ \vec n_a \times \vec n_j] 
\right ),
\label{eq3.37}
\ee
where $\theta_j$ and $\varphi_j$ are the polar and azimuthal angles of the parton $j$. 
The derivation
of Eq.~(\ref{eq3.37}) is provided in Appendix~\ref{app:shift}. It follows that 
\be
\Theta(R_{\xa \yb},q_j) 
= \Theta(R_{jk},q_j)
+\frac{\omega_k}{\sqrt{s}}\; 
R_{jka} \; \delta(R  - R_{jk}) \left ( 
\delta_{q_j,p_j}-
\delta_{q_j,p_{[\xa \yb]}}
\right )
+{\cal O}(\tau^2).
\label{eq3.38}
\ee
The first term on the 
the right-hand side of 
Eq.~(\ref{eq3.38}) is not power-suppressed;  it will have to be combined with corrections to the matrix element, the observable and the phase space.  Therefore, this term  will contribute both at  leading and at next-to-leading order   in the 
expansion in $\tau$.  
  
On the contrary, the  ${\cal O}(\omega_k/\sqrt{s})$ term  in Eq.~(\ref{eq3.38}) is  already 
power-suppressed; it involves two contributions 
with opposite signs which depend on whether the two gluons are clustered into a jet or not. Since this is  a power-suppressed contribution already, the  clustering issue is only relevant for the \emph{jettiness function}, where the difference  between the two cases  in the soft limit is a  \emph{leading order} effect.

Therefore, the power correction
that originates from the expansion of $R_{\xa \yb}$ in Eq.~(\ref{eq3.38}) reads 
\be
\begin{split} 
 &  \frac{{\rm d} \sigma^{s,R}_{\cal O}}{{\rm d} \tau}   =
      g_s^2 \; {\cal N}^{-1} \int  {\rm d} \Phi^{ab}_{j \gamma}
       \; 
[{\rm d} k]  
\;  |{\cal M}_0|^2(p_a,p_b; p_j,p_\gamma) \;  {\cal O}(p_j,  p_\gamma) \; \delta(R - R_{jk})
\\ 
& \times S_1(\vec n_k
)
\; \frac{1}{\omega_k \sqrt{s}}
\;
{\cal R}_{jka} \;
\left ( 
\delta( \tau - \omega_k \psi_{\rm nc}(\vec n_k) ) - 
\delta( \tau - \omega_k \psi_{\rm c}(\vec n_k) ) 
\right ), 
\label{eq3.39}
\end{split}
\ee
where the functions $\psi_{\rm nc,c}$ refer to non-clustered and clustered definitions of the one-jettiness function, respectively. They read 
\be
\begin{split}
& \psi_{\rm nc} =     {\rm min}
   \left \{ \frac{2 E_a \rho_{ak}}{P_a},\frac{2 E_b \rho_{bk}}{P_b},\frac{ 2 E_j \rho_{jk} }{P_J} \right \},\;\;\;\;
   \psi_{\rm c} = \psi_{\rm nc}
   +  \frac{2 E_j \rho_{jk} }{P_J}.
\end{split}
   \ee

Integrating over $\omega_k$ in Eq.~(\ref{eq3.39}), taking the $\ep \to 0$ limit 
and separating the integration over directions of the vector $\vec k$, we obtain 
\be
  \frac{{\rm d} \sigma^{s,R}_{\cal O}}{{\rm d} \tau}   =
       \frac{\alpha_s}{2 \pi} \bar \sigma_0 \int  {\rm d} \Phi^{ab}_{j \gamma}
       \; 
\frac{1- 2\beta + 2\beta^2}{2 \beta (1-\beta)} \; {\cal O}(p_j,  p_\gamma) 
\; {\cal F}_R(\vec n_a, \vec n_b, \vec n_j),
\label{eq3.42a}
\ee
where 
\be
\begin{split}
{\cal F}_R
 = \frac{1}{\sqrt{s}}  \int \frac{{\rm d} \Omega_k}{2\pi} \; 
\delta(R - R_{jk}) \; S_1(\vec n_k
)\;
{\cal R}_{jka}\;
\left ( 
\frac{1}{\psi_{\rm nc}(\vec n_k)} - \frac{1}{\psi_{\rm c}(\vec n_k)}
 \right ).
 \end{split}
 \label{eq3.42b}
\end{equation}
We have mentioned above that, 
when writing Eq.~(\ref{eq3.42a}), we
have taken the $\ep \to 0$ 
limit. To justify this step, we note that the $\delta$-function 
$\delta(R-R_{jk}) $ in 
Eq.~(\ref{eq3.42b})
depends on the polar and azimuthal angles of the gluon $k$, and  the integration over directions of the gluon momentum cannot produce collinear singularities.  To perform it, we integrate first over the
 gluon azimuthal angle $\varphi_k$ and find 
\be
\begin{split}
{\cal F}_R & = 
\frac{R}{2 \pi\sqrt{s}} \; \int \limits_{-1}^{1} 
\frac{{\rm d} \cos \theta_k}{\sqrt{R^2 - (\eta_k - \eta_j)^2 }}
\; \sum \limits_{\alpha=1}^{2}
S_1(\vec n_k^{(\alpha)}
)
\\
& \times {\cal R}_{jka}(\vec n_a, \vec n_k^{(\alpha)}, \vec n_j  )
 \left ( 
\frac{1}{\psi_{\rm nc}(\vec n_k^{(\alpha)})} - \frac{1}{\psi_{\rm c}(\vec n_k^{(\alpha)})}
 \right )
 \theta \left (\pi-\sqrt{R^2 - 
 (\eta_k - \eta_j)^2} \right ).
 \end{split}
\ee
The sum runs over two solutions for the azimuthal angle of the  vector $\vec n_k$.  
The parametrization reads
\be
\vec n_k^{(\alpha)} = (\sin \theta_k \cos \varphi_k^{(\alpha)},
\sin \theta_k \sin \varphi_k^{(\alpha)}, \cos \theta_k),
\ee
where 
\be
\varphi_k^{(\alpha)} = {\rm mod}(\varphi_j \pm \sqrt{R^2 - (\eta_k - \eta_j)^2},2\pi). 
\ee

\paragraph{Remaining contributions}

The remaining soft contributions involve modifications of the matrix element, the phase space and the observables, but not the angular measure of the clustering algorithm. 
Putting everything together, we write the result in the following way
\be
\begin{split}
 \frac{{\rm d} \sigma^{s}_{\cal O}}{{\rm d} \tau}
&= {\cal N}^{-1} [\alpha_s]
\int {\rm d} \Phi^{ab}_{\gamma j}
\;
\frac{{\rm d} \omega_k}{\omega_k^{1+2\ep}} \;
[{\rm d} \Omega_k]
\;
|{\cal M}_0|^2(p_a,p_b;p_j,p_\gamma) \;  \Theta(R_{j k}, q_j)  \;
\delta(\tau - {\cal T}(q_j) )
\\
& \times \Bigg [  
S_1(\vec n_k) \Big  ( 
1+ \frac{\omega_k}{\sqrt{s}} \; \Big  \{ 
2 \ep \;  
- \sum \limits_{x \in j,\gamma}
 \big ( p_x^\mu
+ \frac{\sqrt{s}}{2} \hat{k}^\mu - \frac{\rho_{kx}}{2} P_{ab}^\mu
\big ) \partial_{p_x,\mu}
\\
& + \theta(R-R_{jk}) \sqrt{s}  \hat k^\mu \; 
\partial_{p_j,\mu}
\Big \} 
\Big  )
 {\cal O}(p_j, p_\gamma)
+ \frac{\omega_k}{\sqrt{s} } S_2(\vec n_k)
{\cal O}(p_j, p_\gamma)
\Bigg ],
\end{split}
\ee
where all terms beyond next-to-leading-power corrections in 
$\tau$ 
have been omitted. Similarly  to what has been discussed earlier, 
we integrate over the gluon energy 
$\omega_k$
 using the fact 
that the one-jettiness function is linear in it, c.f. Eq.~(\ref{eq3.39}).  However, instead of employing  functions $\psi_{\rm c,nc}$ introduced earlier, we write 
\be
{\cal T}(q_j) = 
\omega_k \psi_k(\vec n_k),
\ee
which allows us to proceed without indicating whether we deal with the clustered or the unclustered case, until later. 
Integrating over $\omega_k$, 
we find 
\be
\begin{split}
 \frac{{\rm d} \sigma^{s}_{\cal O}}{{\rm d} \tau}
&= {\cal N}^{-1} \frac{ [\alpha_s]}{\tau^{1+2\ep}}
\int {\rm d} \Phi^{ab}_{\gamma j}
[{\rm d} \Omega_k]
\; \psi_{k}^{2\ep}(\vec n_k)
|{\cal M}_0|^2(p_a,p_b;p_j,p_\gamma) \;  \Theta(R_{j k}, q_j) \;
\\
& 
\times \Bigg [  
S_1(\vec n_k) \Big  ( 
1+ \frac{\tau}{\sqrt{s} \;\psi_k(\vec n_k)}\; \Big  \{ 
2 \ep \;  
- \sum \limits_{x \in j,\gamma}
 \big ( p_x^\mu
+ \frac{\sqrt{s}}{2} \hat{k}^\mu - \frac{\rho_{kx}}{2} P_{ab}^\mu
\big ) \partial_{p_x,\mu} 
\\
& + \theta(R-R_{jk}) \sqrt{s}  \hat k^\mu \; 
\partial_{p_j,\mu}
\Big \} 
\Big  )
 {\cal O}(p_j, p_\gamma)
+ \frac{\tau}{\sqrt{s} \;\psi_k(\vec n_k)} S_2(\vec n_k)
{\cal O}(p_j, p_\gamma)
\Bigg ].
\end{split}
\label{eq3.40}
\ee

It remains to integrate Eq.~(\ref{eq3.40}) over 
$\vec n_k$.  This step is non-trivial because the integration 
is divergent, and  the structure of divergences depends on whether the gluons have been clustered in a jet or not.  

Indeed, if the clustering happens, the function $\Theta(R_{jk},q_j)$ restricts the integration over 
the gluon angle to the region around $\vec n_k || \vec n_j$, which implies that this is the only direction that can cause collinear singularities in this case.  On the contrary,  if no clustering occurs, the only possible collinear configurations  are  $\vec n_k || \vec n_a$ and $\vec n_k || \vec n_b$. 

To extract the singularities, and to re-write  the integral in Eq.~(\ref{eq3.40}) in such a way that non-trivial integrations can be performed in three dimensions, 
we employ  the methodology of local subtractions. 
Although  it is fairly straightforward to construct the subtraction terms,  the present case is somewhat special 
 because in  the subleading terms 
 the collinear singularities   
are \emph{power-like}. To see this, we note that in Eq.~(\ref{eq3.40})  functions $S_{1,2}(\vec n_k)$ have  linear singularities in the collinear limits (c.f. Eq.~(\ref{eq3.31})),  leading  to usual logarithmic singularities when the integration over directions of $\vec n_k$  is performed.  However, in 
the subleading terms these singularities are 
further amplified  by singularities caused by the presence 
of the  function $1/\psi_k$ in the integrand. 
 We explain below 
 how suitable  subtraction 
 terms can  be constructed in this case as well. 

\paragraph{The clustered case}

We begin by considering the clustered case where the only 
singular direction is $\vec n_k || \vec n_j$.
To subtract this singularity, we need to expand the integrand in Eq.~(\ref{eq3.40}) around this limit. Since the one-jettiness function, 
as well as the function 
$\theta(R-R_{jk})$ do not change in the   vicinity of the collinear limit, we can replace them with their limiting values, $ \psi_k \to 2 E_j \rho_{jk}/P_J$ and $\theta(R - R_{jk} ) \to 1$,  for the purpose of subtracting \emph{both} leading- and next-to-leading collinear singularities.   It remains to expand $S_{1,2}(\vec n_k)$  through \emph{constant} terms in the $\rho_{jk} \to 0$ limit. 
We find 
\be
\begin{split} 
& S_1(\vec n_k) = 
\frac{2 C_A}{\rho_{kj}}
+ \frac{ C_F}{\beta(1-\beta)}
+\frac{C_A \ep}{1-\ep}
\frac{ (1-2 \beta + 2 \beta^2) }{\beta(1-\beta)} + {\cal O}(\rho_{jk}),
\\
& S_2(\vec n_k) =
-\frac{4 C_A}{\rho_{kj}}
-\frac{C_A \ep }{1-\ep} \; \frac{ (1-2 \beta + 2 \beta^2) }{\beta(1-\beta)} + {\cal O}(\rho_{jk}).
\end{split}
\label{eq3.49}
\ee
To obtain these expressions, we constructed the expansion of $S_{1,2}$ in $\rho_{jk}$, and  integrated the obtained expressions  over components of $\vec n_k$, which are transversal to the direction of 
$\vec n_j$.
While one can perform  this integration in the subtraction term, when the difference of the integrand in Eq.~(\ref{eq3.40}) and the subtraction term is constructed, one needs to keep the  dependence on the transverse components of 
$\vec n_k$ to make sure that all singularities 
of the integrand are removed  \emph{locally}.

Similarly, we need to write the vector $\hat k$ by separating its component along $\vec n_j$ from  the transversal ones.  We find 
\be
\frac{\sqrt{s}}{2} \hat k^\mu = (1 - \rho_{jk}) p_j^\mu + \frac{\rho_{jk}}{2} P_{ab}^\mu + \frac{\sqrt{s \rho_{jk} (2-\rho_{jk})}}{2} {\hat k}_\perp^\mu,
\ee
where ${\hat k}_\perp^\mu$ is a four-dimensional unit vector 
( $\hat k_{\perp}^\mu \hat k_{\perp,\mu} = -1 $), 
orthogonal to $\vec n_j$.  Using this decomposition, we obtain 
\be
\begin{split} 
 &S_1  (\vec n_k)
 \Big (
\sum \limits_{x \in j,\gamma}
 \big ( p_x^\mu
+ \frac{\sqrt{s}}{2} \hat{k}^\mu - \frac{\rho_{kx}}{2} P_{ab}^\mu
\big ) \partial_{p_x,\mu} 
- \sqrt{s}  \hat k^\mu \; 
\partial_{p_j,\mu}
\Big  )
     = \\
     & -  C_A
      \Bigg\{ 2 p_\gamma^\mu 
      - \frac{1-2\beta}{2(1-\ep)} 
      \left[\frac{p_a^\mu}{\beta} - \frac{p_b^\mu}{(1-\beta)} -  \frac{(1-2\beta)}{\beta(1-\beta)} \; p_\gamma^\mu \right] \Bigg\}
      \left( \partial_{p_j,\mu} - \partial_{p_\gamma,\mu} \right)+ {\cal O}(\rho_{jk}),
\end{split}
\label{eq3.51}
\ee
where we have expanded $S_1(\vec n_k)$ around the $\vec n_k || \vec n_j $ limit and have averaged   over directions of $k^\mu_\perp$.

We use Eqs~(\ref{eq3.49},\ref{eq3.51})  to write  the linear power correction, that originates from the clustered case,  in the following way
\be
\begin{split}
 \frac{ {\rm d } \sigma_{\cal O}^{s, \rm NLP}}{{\rm d} \tau}
\Big |_{\rm cl}
= & \frac{{\cal N}^{-1} [\alpha_s]}{\sqrt{s} \tau^{2\ep}}
\left ( \frac{P_J} {2 \sqrt{s}}
\right )^{1-2 \ep}
{\rm d} \Phi_{\gamma j}^{ab} \;
| {\cal M}_0(p_a,p_b,p_j,p_\gamma)
|^2 \\
& \times \left ( S_c + 
\int [{\rm d} \Omega_k]
\; 
{\cal F}_k^{\rm cl}
\right ) {\cal O}(p_j, p_\gamma).
\end{split}
\label{eq3.41}
\ee
In Eq.~(\ref{eq3.41}) $S_c$ is the integrated 
subtraction term given by 
\be
\begin{split} 
S_c = & 
\frac{2 C_F}{\beta(1-\beta)} - \frac{C_A (1-2 \beta + 2 \beta^2)}{\beta (1-\beta) }
\\
& +\frac{C_A}{\ep}
      \Bigg\{ 2 p_\gamma^\mu 
      - \frac{1-2\beta}{2(1-\ep)} 
      \left[\frac{p_a^\mu}{\beta} - \frac{p_b^\mu}{(1-\beta)} -  \frac{(1-2\beta)}{\beta(1-\beta)} \; p_\gamma^\mu \right] \Bigg\}
      \left( \partial_{p_j,\mu} - \partial_{p_\gamma,\mu} \right),
\end{split}
\ee
and 
\be
\begin{split} 
& {\cal F}_k^{\rm cl} = 
\frac{2 \sqrt{s} \; \theta(R-R_{jk})}{P_J \; \psi_k}
\Bigg [ S_2(\vec n_k) \\
& - S_1(\vec n_k)
\left ( \big ( p_j^\mu
- \frac{\sqrt{s}}{2} \hat{k}^\mu - \frac{\rho_{jk}}{2} P_{ab}^\mu
\big ) \partial_{p_j,\mu} + \big ( p_\gamma^\mu
+ \frac{\sqrt{s}}{2} \hat{k}^\mu - \frac{\rho_{\gamma k}}{2} P_{ab}^\mu
\big ) \partial_{p_\gamma,\mu} \right ) \Bigg ]
\\
& -\frac{C_A}{\rho_{jk}}
\Bigg [ -\frac{4}{\rho_{jk}}
+ \frac{(1-2\beta)}{2 \beta (1-\beta) } \frac{\sqrt{2} }{\sqrt{\rho_{jk}}} ( \vec n_{k,\perp} \cdot \vec n_a
)
+ \frac{(1-2 \beta +2 \beta^2)}{ \beta (1-\beta)}
\left (1 - \frac{ (\vec n_{k, \perp} \cdot \vec n_a)^2}{2 \beta(1-\beta) } \right ) \Bigg ] 
\\
& -  \frac{C_A }{\rho_{jk}}
      \Bigg[ 2 p_\gamma^\mu 
      - \frac{(1-2\beta)}{2 \beta (1-\beta)} (\vec n_{k, \perp} \cdot \vec n_a) \sqrt{s} \; \hat k_\perp^\mu
       + \frac{\sqrt{2s}}{\sqrt{\rho_{jk}}} \hat k_\perp^\mu \Bigg]
      \left( \partial_{p_j,\mu} - \partial_{p_\gamma,\mu} \right).
\end{split}
\label{eq3.46}
\ee
The transverse vector $\vec n_{k,\perp}$, which parametrizes spatial components of the four-vector $k_\perp$, 
is defined by the following  equation
\be
\vec n_k = (1-\rho_{jk}) \vec n_j
+ \sqrt{\rho_{jk}(2-\rho_{jk})} \; \vec n_{k,\perp},
\ee
with $\vec n_{k,\perp} \cdot \vec n_j = 0$.
Integration over directions  
of the vector $\vec n_k$ in Eq.~(\ref{eq3.41}) is finite and, therefore, is performed 
in the three-dimensional space.

\paragraph{The non-clustered case.}

We continue with  the non-clustered case, where the singularities arise if  $\vec n_k || \vec n_a$ or  $\vec n_k || \vec n_b$.  We 
construct the expansion of the integrand in Eq.~(\ref{eq3.40}) around these limits, closely following  what has been done 
in the clustered case.  
To this end, we again  replace $\theta(R-R_{jk}) \to 1$, 
and $\psi_k(\vec n_k) 
\to 2 E_x \rho_{xk}/ P_x$ where  
$x \in (a,b)$, depending on the collinear limit that we consider.

We first construct the  expansion of the integrand for the  $\vec n_k || \vec n_a$ case. The expansion of the functions $S_{1,2}$ for $\vec n_k || \vec n_a$ reads 
\be
\begin{split} 
& S_1(\vec n_k)= 
\frac{2 C_F}{\rho_{ak}}
+  C_F
+\frac{C_A }{1-\ep}\;
\frac{ \beta  }{1-\beta} + {\cal O}(\rho_{ak}),
\\
& S_2(\vec n_k)  =
-\frac{C_A  }{1-\ep} \; \frac{ (1 -\ep+\beta) }{1-\beta} + {\cal O}(\rho_{ak}),
\\
\end{split}
\ee
where we have  averaged  over directions of the  vector $\vec n_k$ which are orthogonal to $\vec n_a$. It is peculiar that $S_2(\vec n_k)$ does not have the corresponding collinear singularity. 

We also need to re-write the   vector $\hat k$ that 
appears in Eq.~(\ref{eq3.40}), separating its components 
orthogonal to the collision axis,  and averaging over their directions. To this end, we write 
\be
\frac{\sqrt{s}}{2} \hat k^\mu = (1 - \rho_{ak}) p_a^\mu + \frac{\rho_{ak}}{2} P_{ab}^\mu + \frac{\sqrt{s \rho_{ak} (2-\rho_{ak})}}{2} k_\perp^\mu,
\ee
where $k_\perp^\mu$ is related to the spatial direction perpendicular to $\vec n_a$. With this we get

\begin{align} 
 &S_1 (\vec n_k) 
\sum \limits_{x \in j,\gamma}
 \big ( p_x^\mu
+ \frac{\sqrt{s}}{2} \hat{k}^\mu - \frac{\rho_{kx}}{2} P_{ab}^\mu
\big ) \; \partial_{p_x, \mu}
     = \nonumber\\
    & \frac{2 C_F}{\rho_{ak}} \left[ \left( p_j^\mu + \beta p_a^\mu - \bar \beta  p_b^\mu \right) \partial_{p_j\mu} + \left( p_\gamma^\mu + \bar \beta  p_a^\mu - \beta p_b^\mu \right) \partial_{p_\gamma\mu} \right]  \\ 
    & + \Bigg\{\Bigg[\frac{C_A}{2 (1-\ep) \bar \beta }
      \left(p_j^\mu (1+2\beta) + \beta p_a^\mu - \bar \beta  p_b^\mu \right)  + C_F\left(p_j^\mu -\beta p_a^\mu + \bar \beta p_b^\mu      \right) \Bigg]\partial_{p_j\mu}  \nonumber\\
      & +\Bigg [C_F\left(p_\gamma^\mu -\bar \beta  p_a^\mu + \beta p_b^\mu    \right) - \frac{C_A}{2 (1-\ep) \bar \beta }\left(p_\gamma^\mu (1-2\beta) - \bar \beta  p_a^\mu + \beta p_b^\mu \right) \Bigg]\partial_{p_\gamma\mu} \Bigg\} + {\cal O}(\rho_{ak}) \nonumber,
\end{align}
where we introduced the short-hand notation $\bar \beta = 1-\beta$.

The last potential  collinear singularity to consider is  $\vec n_k || \vec n_b$.  The analysis in this case  follows  steps 
discussed in connection with $\vec n_k || \vec n_j$ and $\vec n_k || \vec n_a$ singularities. We find 
\be
\begin{split} 
& S_1(\vec n_k) = 
\frac{2 C_F}{\rho_{bk}}
+  C_F
+\frac{C_A }{1-\ep}\;
\frac{ 1- \beta  }{\beta} + {\cal O}(\rho_{bk}),
\\
& S_2(\vec n_k) = 
-\frac{C_A  }{1-\ep} \; \frac{ (2-\beta -\ep) }{\beta} + {\cal O}(\rho_{bk}),
\end{split}
\ee
where we have averaged over directions of the vector $k$ orthogonal to the collision axis.  Writing 
momentum $\hat k$ in terms of $p_b$ and the transverse component, 
\be
\frac{\sqrt{s}}{2} \hat k^\mu = (1 - \rho_{bk}) p_b^\mu + \frac{\rho_{bk}}{2} P_{ab}^\mu + \frac{\sqrt{s \rho_{bk} (2-\rho_{bk})}}{2} k_\perp^\mu,
\ee
and averaging over directions of $k_\perp$,  
we find 
\begin{align} 
 & S_1 (\vec n_k)
\sum \limits_{x \in j,\gamma}
 \big ( p_x^\mu
+ \frac{\sqrt{s}}{2} \hat{k}^\mu - \frac{\rho_{kx}}{2} P_{ab}^\mu
\big ) \; \partial_{p_x,\mu}
      = \nonumber\\
    & \frac{2 C_F}{\rho_{bk}} \left[ \left( p_j^\mu - \beta p_a^\mu + \bar \beta  p_b^\mu \right) \partial_{p_j,\mu} + \left( p_\gamma^\mu - \bar \beta  p_a^\mu + \beta p_b^\mu \right) \partial_{p_\gamma,\mu} \right]  \\ 
    & +\Bigg\{\Bigg[\frac{C_A}{2 (1-\ep) \beta}
      \left(p_j^\mu (3-2\beta) - \beta p_a^\mu + \bar \beta p_b^\mu \right)  + C_F\left(\beta p_a^\mu - \bar \beta p_b^\mu + p_j^\mu    \right) \Bigg]\partial_{p_j,\mu} \nonumber \\
      & +\Bigg [C_F\left(
      \bar \beta  p_a^\mu - \beta p_b^\mu + p_\gamma^\mu    \right) + \frac{C_A}{2 (1-\ep) \beta}\left(p_\gamma^\mu (1+2\beta) - \bar \beta  p_a^\mu + \beta p_b^\mu \right) \Bigg]\partial_{p_\gamma,\mu} \Bigg\} + {\cal O}(\rho_{bk}). \nonumber
\end{align}

We use the above results  to write  the linear power correction to the non-clustered contribution in the following way
\be
\begin{split}
     \frac{ {\rm d } \sigma_{\cal O}^{s, \rm NLP}}{{\rm d} \tau}
\Big |_{\rm nc}
& = \frac{{\cal N}^{-1} [\alpha_s]}{\sqrt{s} \tau^{2\ep}}
\left ( \frac{1} {\sqrt{s}}
\right )^{1-2 \ep}
{\rm d} \Phi_{\gamma j}^{ab} \;
| {\cal M}_0(p_a,p_b,p_j,p_\gamma)
|^2 \\
& \times \left ( P_a^{1-2\ep} S_{ ca} + P_b^{1-2\ep} S_{ cb} +
\int [{\rm d} \Omega_k] \;
{\cal F}_k^{\rm nc} 
\right ) {\cal O}(p_j, p_\gamma).
\end{split}
\label{eq3.42}
\ee
The terms  $S_{ca}$
and $S_{cb}$ describe the integrated 
subtraction term for the limit $\vec n_k || \vec n_a$ and $\vec n_k || \vec n_b$ respectively; 
all the $1/\ep$ divergences are collected there. 
These terms  read
\be
\begin{split} 
& S_{ca} = 2 C_F  + C_A \frac{ \beta}{\bar \beta } - \frac{C_A }{\ep} \frac{(1+\beta)}{ \bar \beta}
\\
& -\frac{1}{\ep}  \Bigg\{\Bigg[\frac{C_A}{2 (1-\ep) \bar \beta}
      \left(p_j^\mu (1+2\beta) + \beta p_a^\mu - \bar \beta p_b^\mu \right)  + C_F\left(p_j^\mu -\beta p_a^\mu + \bar \beta p_b^\mu      \right) \Bigg]\partial_{p_j,\mu}  \\
      & +\Bigg [C_F\left(p_\gamma^\mu -\bar \beta  p_a^\mu + \beta p_b^\mu    \right) - \frac{C_A}{2 (1-\ep) \bar \beta }\left(p_\gamma^\mu (1-2\beta) - \bar \beta  p_a^\mu + \beta p_b^\mu \right) \Bigg]\partial_{p_\gamma,\mu} \Bigg\} ,
\end{split}
\ee
and 
\be
\begin{split} 
& S_{cb} = 
 2 C_F + C_A\frac{\bar \beta}{\beta } - \frac{C_A}{\ep} \frac{(2-\beta)}{\beta}
\\
& -\frac{1}{\ep}\Bigg\{\Bigg[\frac{C_A}{2 (1-\ep) \beta}
      \left(p_j^\mu (3-2\beta) - \beta p_a^\mu + \bar \beta p_b^\mu \right)  + C_F\left(p_j^\mu + \beta p_a^\mu - \bar \beta p_b^\mu   \right) \Bigg]\partial_{p_j,\mu}  \\
      & +\Bigg [C_F\left(p_\gamma^\mu + \bar \beta p_a^\mu - \beta p_b^\mu \right) + \frac{C_A}{2 (1-\ep) \beta}\left(p_\gamma^\mu (1-2\beta) - \bar \beta p_a^\mu + \beta p_b^\mu \right) \Bigg]\partial_{p_\gamma,\mu} \Bigg\} .
\end{split}
\ee
The remaining integral over directions of the vector $\vec k$ in  Eq.~(\ref{eq3.42}) is finite, and can be performed in three dimensions.  The function ${\cal F}_k^{\rm nc}$ reads 
\be
\begin{split} 
& {\cal F}_k^{\rm nc} = 
\frac{ \sqrt{s} \; \theta(R_{jk}-R)}{\; \psi_k}
\Bigg [ S_2(\vec n_k) - S_1(\vec n_k)
\sum \limits_{x \in j,\gamma} 
(p_x^\mu + \frac{\sqrt{s}}{2} \hat k^\mu - \frac{\rho_{xk}}{2} P_{ab}^\mu ) \; \partial_{p_x,\mu}
\Bigg ] \\
&-\Bigg \{\frac{P_a C_A}{\rho_{ak}}
\Bigg [ 
- \frac{1}{2   \bar \beta } \frac{\sqrt{2} }{\sqrt{\rho_{ak}}} ( \vec n_{k,\perp a} \cdot \vec n_j
)
-\frac{1}{   \bar \beta}
\left (1 +  \frac{ (\vec n_{k, \perp a} \cdot \vec n_j)^2}{2  \bar \beta } \right ) \Bigg ] 
\\
&+\frac{P_b C_A}{\rho_{bk}}
\Bigg [ 
- \frac{1}{2 \beta } \frac{\sqrt{2} }{\sqrt{\rho_{bk}}} ( \vec n_{k,\perp b} \cdot \vec n_j
)
- \frac{1}{ \beta}
\left (1 + \frac{ (\vec n_{k, \perp b} \cdot \vec n_j)^2}{2 \beta } \right ) \Bigg ] \Bigg\} \\
& + \frac{P_a}{\rho_{ak}}\Bigg\{ S_{a,p_j}^{\mu}\partial_{p_j,\mu}  + S_{a,p_\gamma}^{\mu} \partial_{p_\gamma,\mu}  \Bigg\} + \frac{P_b}{\rho_{bk}}\Bigg\{ S_{a,p_j}^{\mu}\partial_{p_j,\mu}  + S_{a,p_\gamma}^{\mu} \partial_{p_\gamma,\mu} 
\Bigg \}_{
\;  \substack{\beta \leftrightarrow \bar \beta \\  \; p_a \leftrightarrow p_b}},
\end{split}
\label{eq3.57}
\ee
where 
\begin{align} 
    S_{a,p_j}^{\mu} & =  \frac{C_A}{2\bar \beta}
      \bigg [ (p_j^\mu +  p_a^\mu)\frac{(\vec n_{k,\perp a} \cdot \vec n_j)^2}{ \bar\beta}  + (\vec n_{k,\perp a} \cdot \vec n_j)\sqrt{s} \hat k_{\perp }^\mu 
      \nonumber \\
      & + \frac{\sqrt{2} \; (\vec n_{k,\perp a} \cdot \vec n_j) }{ \sqrt{\rho_{ak}}}\left(p_j^\mu + \beta p_a^\mu - \bar \beta p_b^\mu \right) \bigg ] + C_F \Bigg[ 2 \;\frac{p_j^\mu + \beta p_a^\mu - \bar \beta p_b^\mu }{\rho_{ak}}  \\
      &  + p_j^\mu -\beta p_a^\mu + \bar \beta p_b^\mu + \frac{\sqrt{2}}{ \sqrt{\rho_{ak}}} \left( (\vec n_{k,\perp a} \cdot \vec n_j) P_{ab}^\mu + \sqrt{s} \; \hat k_{\perp }^{\mu}\right)  \Bigg ], \nonumber  \\
  S_{a, p_\gamma}^{\mu} &= - \frac{C_A}{2\bar \beta}\bigg[(p_b^\mu -  p_\gamma^\mu)\frac{(\vec n_{k,\perp a} \cdot \vec n_j)^2}{ \bar\beta}  - (\vec n_{k,\perp a} \cdot \vec n_j)\sqrt{s} \hat k_{\perp }^\mu  \nonumber \\
  & - \frac{\sqrt{2} \; (\vec n_{k,\perp a} \cdot \vec n_j) }{\sqrt{\rho_{ak}}}\left(p_\gamma^\mu + \bar \beta p_a^\mu - \beta p_b^\mu \right)\bigg ] +C_F\bigg [ 2 \;\frac{p_\gamma^\mu + \bar \beta p_a^\mu - \beta p_b^\mu }{\rho_{ak}}    \\
  & + p_\gamma^\mu -\bar \beta p_a^\mu + \beta p_b^\mu + \frac{\sqrt{2}}{\sqrt{\rho_{ak}}} \left( - (\vec n_{k,\perp a} \cdot \vec n_j) P_{ab}^\mu + \sqrt{s} \; \hat k_{\perp }^{\mu}\right) \bigg ] \nonumber .
\end{align}
In the above equations we again have used $\bar \beta = 1- \beta$, and we introduced the unit  transverse vectors $\vec n_{k,\perp a}$, $\vec n_{k,\perp b}$ 
defined as follows
\be
\begin{split}
&\vec n_{kx} = (1-\rho_{xk}) \vec n_x
+ \sqrt{\rho_{xk}(2-\rho_{xk})} \; \vec n_{k,\perp x},
\;\;\; x \in (a,b).
\end{split}
\ee
This completes our discussion of the soft contribution to 
next-to-leading-power corrections. The final result is obtained as a sum of Eqs~(\ref{eq3.42a},\ref{eq3.41},\ref{eq3.42}). These results still contain $1/\ep$ poles which, however, are only present in the  integrated subtraction terms.  
These $1/\ep$ poles cancel against the ones from the collinear contributions to power corrections that we will now discuss.

\subsection{The case  $\yb || a$ }

We consider the collinear  
$\yb || a$ case.
Our starting point is Eq.~(\ref{eq2.15a})
which we repeat here for convenience 
\be 
  \frac{{\rm d} \sigma^{ca}_{\cal O}}{{\rm d} \tau}  
  = {\cal N}^{-1}\int  {\rm d} \Phi(p_a,p_b|\tilde p_j,p_\yb,\tilde p_\gamma) |{\cal M}|^2(p_a,p_b;\tilde p_j,p_\yb, \tilde p_\gamma)
  \delta \left ( \tau - \frac{2 p_a p_\yb}{P_a} \right ) {\cal O}(\tilde p_j, \tilde p_\gamma). 
\ee
We note that we renamed $p_\xa \to \tilde p_j$ for reasons explained earlier. 

To extract the subleading one-jettiness 
contribution,  we proceed in the same way as  in the case of the 
color-singlet production~\cite{Agarwal:2025dvo}. To align  the current notations with   that reference, we will re-name $p_\yb \to k$.   Following the discussion in 
Ref.~\cite{Agarwal:2025dvo}, we decompose the momentum $k$ as
\be
k = (1-x) p_a + \tilde k_a,
\label{eq3.71}
\ee
where $(1-x) = k \cdot P_{ab}/p_a \cdot p_b$.  The momentum 
conservation becomes 
\be
x p_a + p_b =
\tilde p_j + \tilde p_\gamma + \tilde k_a.
\ee
Since the invariant masses of 
vectors 
$\tilde Q = \tilde p_j + \tilde p_\gamma$ 
and  $ \tilde Q+ \tilde k_a$ are the same \cite{Agarwal:2025dvo}, one can obtain the latter by performing the Lorentz boost of the former. We denote the required Lorentz boost by $\Lambda_a$,\footnote{This matrix is given explicitly  in 
Appendix~A of Ref.~\cite{Agarwal:2025dvo}.
}
and write
\be
\tilde p_j + \tilde p_\gamma + \tilde k_a = p_j + p_\gamma, 
\ee
with  
\be
p_j = \Lambda_a \tilde p_j,
\;\;\;\;
p_\gamma = 
\Lambda_a \tilde p_\gamma. 
\ee

Using the boost and the   parametrization of the phase space  from Ref.~\cite{Agarwal:2025dvo}, we find
\be
\begin{split} 
 \frac{{\rm d} \sigma^{ca} }{{\rm d} \tau}
& = \frac{C_F [\alpha_s] P_a^{1-\ep} }{2 \tau^{1+\ep}} 
{\cal N}^{-1} \;
\int \limits_0^1 {\rm d} x \; {\rm d} \Phi^{xa,b}_{\gamma j} 
\;  \left[ {\rm d} \Omega_n^{(d-2)} \right]  \;   (1-x)^{-\ep}  \left ( 1 + \frac{ \ep \rho^*_{ak}}{2}  \right )
\\
& \times 
       \; {\cal O}( \Lambda_a^{-1}  p_j, \Lambda_a^{-1} p_\gamma)
       \; \sum \limits_{\rm pol, col}^{}  C_F^{-1} g_s^{-2} \tau \ |{\cal M}(p_b,p_a,k,\Lambda_a^{-1}  p_j, \Lambda_a^{-1} p_\gamma ) |^2.
\end{split}
\label{eq3.58}
\ee
In Eq.~(\ref{eq3.58})  ${\rm d} \Phi_{\gamma j}^{xa, b}$ denotes the  phase space of  partons with momenta $p_{j,\gamma}$,  produced in a collision of a parton with momentum $x p_a$ and $p_b$, 
and 
\be
\rho_{ak}^* = \frac{2 P_a \tau}{s(1-x)}.
\ee

The boost matrix $\Lambda_a$ depends on the four-vector  $\tilde k_a$, 
which can be parametrized as 
\be
\tilde k_a = \frac{2 k p_a}{s} (p_b - p_a) 
+ k_{\perp,a}.
\ee
The vector  $k_{\perp,a}$ is orthogonal to $p_{a,b}$.
Since the  emission angle of the gluon with the momentum $k$ relative to the collision axis scales as $\theta \sim \sqrt{\tau/\sqrt{s}}$, the transverse momentum 
$k_{\perp,a}$ scales as $k_{\perp,a} \sim \sqrt{\tau}$. This implies that $k p_a \sim \tau$.

As  follows from Eq.~\eqref{eq3.58}, to compute the power corrections, we need to expand  both the matrix element and the observable in $\tau$. Since  $ \Lambda_a^{-1}  p_j$ and $ \Lambda_a^{-1}  p_\gamma$  deviate from $p_{j,\gamma}$ by terms proportional to $\tilde k_a  \sim \sqrt{\tau}$, we need to expand the observable ${\cal O}$ up to the  second order
in $\tilde k_a$. 
Furthermore, the expansion around collinear limits introduces soft 
$(x \to 1)$ singularities in the expansion terms.  These singularities need to be extracted,  and we discuss below how we deal with this problem. 

We note in this respect that in the current paper we work with  the particular   matrix element squared,  and we do not attempt to repeat 
a more general approach described in Ref.~\cite{Agarwal:2025dvo} for colorless final states. 
Hence, we use the explicit form of the matrix element squared of the process  $ \bar q + q \to gg \gamma$, and the explicit expression for the boost, to construct 
the 
expansion of the matrix element squared and the observable through next-to-leading power. Collecting terms  that become singular in the  $x \rightarrow 1$ limit, we find 
\be
\begin{split} 
 \frac{{\rm d} \sigma^{ca,{\rm NLP}} }{{\rm d} \tau} \Big |_{ x \to 1}
 & = \frac{ [\alpha_s]  }{ \tau^{1+\ep}} \; \frac{P_a^{1-\ep} \tau}{2s}
  \tilde \sigma_0 
  \\
  & \times \int \limits_0^1 \; {\rm d} x  \; {\rm d} \Phi^{xa,b}_{\gamma j}   
  \; \frac{( 1 - 2\beta + 2 \beta^2 - \ep )}{\beta \bar \beta}
\; (1-x)^{-1-\ep} \;
\\
& \times \Bigg  \{
  \frac{ 2(1+\ep)C_F }{(1-x)}   - \frac{(1+\beta-\ep)C_A}{(1-\ep) \bar \beta } + \frac{2\beta C_A}{ (1-x) (1-\ep) \bar \beta} \\
  & -  C_A\left[\frac{\beta(1-2\beta)p_a^\mu - \bar \beta (1-2\beta) p_b^\mu +p_j^\mu}{2 \bar \beta (1-\ep)}\right] \partial_{p_j, \mu} \\
   & +  C_A\left[\frac{\beta(1-2\beta)p_b^\mu - \bar \beta (1-2\beta) p_a^\mu +p_\gamma^\mu}{2 \bar \beta (1-\ep)}\right] \partial_{p_\gamma, \mu}\\
    & + 2 C_F \left[  \left (  \beta  p_a^\mu - \bar \beta p_b^\mu \right )\partial_{p_j, \mu} + \left ( \bar \beta p_a^\mu - \beta p_b^\mu \right ) \partial_{p_\gamma, \mu}    \right ] 
  \Bigg  \} \; {\cal O}(p_j,p_\gamma).
  \label{eq3.53}
\end{split} 
\ee
We note that the parameter   $\beta$  in this case refers to the Sudakov parametrization of momenta $p_{j,\gamma}$, and $\bar \beta = 1-\beta$. The required parametrization can be obtained from   Eq.~(\ref{eq4.4}) provided that one replaces there  $p_a \to x p_a$ and $s \to xs $. The same applies to  the phase space ${\rm d} \Phi_{\gamma j}^{xa, b}$ -- one can use Eq.~(\ref{eq3.7})  provided that $s$ is replaced with $x s$ there.

It follows from Eq.~(\ref{eq3.53}) that there is a logarithmic and a power-like singularity in the
term that contains an observable  ${\cal O}(p_j,p_\gamma)$, and a logarithmic singularity  in the terms  with derivatives of the observable   ${\cal O}$. 
The logarithmic singularities are standard; we deal with them by expressing  $(1-x)^{-1-\ep}$ in Eq.~(\ref{eq3.53}) through $\delta(1-x)$ 
and plus-distributions. 

On the contrary, power-like singularities are unusual, and the easiest way to deal with them  is to integrate by parts.  We write  
\be
\int \limits_{0}^{1} 
{\rm d} x \; \frac{F(x)}{(1-x)^{2+\ep}} 
= -\frac{1}{1+\ep}
\int \limits_{0}^{1} 
{\rm d} x \; 
(1-x)^{-1-\ep} 
\; \frac{\partial F }{\partial x }, 
\ee
where we made use of the fact that the  boundary term 
at $x = 0$ drops out because 
it corresponds to a collision where the parton $a$ has a vanishing  four-momentum. 

In the context of Eq.~(\ref{eq3.53}), the  function $F(x)$ is a product of the phase-space element that contains  the factor $x^{-\ep}$, and the observable ${\cal O}$. In fact, ${\cal O}$ is the only quantity where computation of the derivative with respect to  $x$ requires further discussion. The dependence of the observable ${\cal O}$ 
on $x$ arises through the dependences of $p_j$ and $p_\gamma$ on this variable. We find 
\be
\begin{split}
\partial_x  {\cal O}(p_j,p_\gamma) 
& = \frac{1}{2} 
\left ( 
\beta p_a^\mu + 
\frac{1}{x} (p_{j}^\mu - \bar \beta  p_b^\mu)
\right )
\partial_{p_{j,\mu}} {\cal O}
+\frac{1}{2} 
\left ( 
\bar \beta  p_a^\mu + 
\frac{1}{x} (p_{\gamma}^\mu - \beta p_b^\mu)
\right )
\partial_{p_{\gamma,\mu}} {\cal O}.
\end{split}
\ee
Since at this point all the divergences are logarithmic, it is straightforward to extract them by rewriting $1/(1-x)^{1+\ep}$ through the plus-distributions and the function  $\delta(1-x)$.  
Putting everything together, we find that the divergent contribution reads 
\be
\begin{split} 
& \frac{{\rm d} \sigma^{ca,{\rm NLP}} }{{\rm d} \tau} \Big |_{\rm div}
  = \frac{ [\alpha_s]  }{ \ep} \; \frac{P_a }{2s }\left(\tau P_a\right)^{-\ep}
  \tilde \sigma_0 \; {\rm d} \Phi^{a b}_{\gamma j}   \; \frac{(1 - 2\beta +2\beta^2 -\ep)}{\beta \bar \beta}  \;
\\
&  \times \Bigg  \{ C_F \Bigg[\left( p_j^\mu -\beta p_a^\mu + \bar \beta p_b^\mu      \right) \partial_{p_j,\mu}  +\left( p_\gamma^\mu -\bar \beta p_a^\mu + \beta p_b^\mu    \right)  \partial_{p_\gamma,\mu} \Bigg]\\
& 
 + C_A \Bigg[\frac{(1+\beta)}{\bar \beta} +\frac{1}{2\bar \beta}
      \left(p_j^\mu (1+2\beta) + \beta p_a^\mu - \bar \beta p_b^\mu \right)  \partial_{p_j, \mu}  \\
  & \;\;\;\;\;- \frac{1}{2 \bar \beta}\left(p_\gamma^\mu (1-2\beta) - \bar \beta p_a^\mu + \beta p_b^\mu \right) \partial_{p_\gamma, \mu}  \Bigg] 
  \Bigg  \} \; {\cal O}(p_j,p_\gamma). 
  \label{eq3.66}
\end{split} 
\ee
We note that all momenta in the above expression are evaluated at  $x = 1$; this is indicated, in particular, by  the fact that it contains the phase-space element ${\rm d} \Phi^{ab}_{\gamma j}$.
The finite contribution to the NLP cross section evaluates to 

\begin{align}
    &\frac{{\rm d} \sigma^{ca,{\rm NLP}} }{{\rm d} \tau} \Big |_{\rm fin} = \frac{[\alpha_s] P_a }{s} \; \tilde \sigma_0 \; \frac{(1 - 2\beta + 2 \beta^2)}{2\beta \bar \beta} 
    \nonumber 
    \\
     &\times \int \limits_0^1  {\rm d} x \; {\rm d} \Phi^{xa, b}_{\gamma j} \;  \bigg \{  - 4 \delta(1-x) \left(C_F + \frac{3 C_A \beta}{4 \bar \beta} \right)  - C_A \; \mathcal{L}_0 (1-x) \left( \frac{1+\beta}{\bar \beta} \right) 
     \nonumber \\
     &   + C_F \; \mathcal{L}_0 (1-x)   \bigg [ (\bar \beta p_a ^\mu 
      -  \beta p_b ^\mu - p_\gamma ^\mu ) \partial_{p_\gamma,\mu}
       + ( \beta p_a^\mu - \bar \beta p_b ^\mu - p_j ^\mu)\partial_{p_j,\mu} \bigg ] \nonumber \\
     &  + C_A \;\frac{\delta (1-x) }{2 \bar \beta} \bigg [  ( \bar \beta (1- 2\beta  )p_a ^\mu -\beta (1-2\beta) p_b ^\mu  - {p_\gamma}^\mu) \partial_{p_\gamma,\mu} 
     \label{eq3.70}
     \\
     &+(  \beta (1- 2\beta  )p_a ^\mu -\bar \beta (1-2\beta) p_b ^\mu  + {p_j}^\mu )\partial_{p_j,\mu} \bigg] 
     \nonumber \\
     &  + C_A \;\frac{\mathcal{L}_0 (1-x) }{2 \bar \beta} \bigg [  ( \beta  p_b ^\mu  - \bar \beta p_a ^\mu + (1 - 2\beta){p_\gamma}^\mu) \partial_{p_\gamma,\mu} 
     \nonumber \\
     & + ( \bar \beta  p_b ^\mu  - \beta p_a^\mu - (1 + 2\beta){p_j}^\mu) \partial_{p_j,\mu}\bigg] + \frac{\beta \bar \beta}{4(1 - 2\beta + 2 \beta^2)} \; R_{\rm ca} (\beta,x, p_a, p_b ) \bigg \} \; {\cal O}(p_j,p_\gamma),
\nonumber 
\end{align}
where ${\cal L}_0(1-x) = 1/(1-x)_+$ and $R_{\rm ca}$ is given by the following expression 
\begin{align}
    R_{\rm ca}&(\beta,x, p_a, p_b ) = \;  \frac{C_F}{\beta \bar\beta}\Bigg [ \frac{16\beta^4 -32\beta^3 +18\beta^2 -2\beta +5}{\beta \bar \beta} \left( 1+ \frac{1}{x^2} \right)  
    \nonumber  \\
    & + \frac{4(2\beta^4 -4\beta^3 + 3\beta^2 -\beta -2)}{\beta \bar \beta x}+ g_2(x, \beta) \; p_\gamma^\mu \partial_{p_\gamma, \mu} 
    + g_2(x,\bar \beta)\; p_j^\mu \partial_{p_j,\mu} \nonumber \\ 
 &+g_1(x,\bar \beta)  \; p_a^\mu \partial_{p_\gamma, \mu}   + \frac{g_1\left (x_1,\bar \beta \right ) }{x}  \;   p_b^\mu \partial_{p_j, \mu} 
  \nonumber + g_1(x, \beta) p_a^\mu \partial_{p_j, \mu}  + \frac{g_1\left (x_1,\beta \right ) }{x}  p_b^\mu \partial_{p_\gamma, \mu} 
 \nonumber \\ 
&+ 4 f_0(\beta) p_a^\mu \left( \bar \beta \partial_{p_\gamma, \mu} + \beta \partial_{p_j, \mu} \right) + \frac{(1+x^2) f_0(\beta)}{2 x^2} \bigg\{ \Big[ -2 \bar \beta^2 x^2 \ p_a^\mu p_a^\nu - 2 \beta^2  \ p_b^\mu p_b^\nu 
 \nonumber \\ 
& - x \ (p_a p_b) \ g^{\mu \nu} + 2 \left( x \ p_a^\mu p_\gamma^\nu + p_b^\mu p_\gamma^\nu \right)  + 4 x \beta \bar \beta \ p_a^\mu p_b^\nu \Big]\partial_{p_\gamma,\nu} \partial_{p_\gamma, \mu} 
 \nonumber \\
& + \Big[ \left(f_0  (\beta) - 2 \right) \left( x^2 \ p_a^\mu p_a^\nu + p_b^\mu p_b^{\nu}  \right) + \left( x \ p_\gamma^\mu p_a^\nu + p_\gamma^\mu p_b^\nu \right) + \left( x \ p_a^\mu p_j^\nu + p_b^\mu p_j^\nu \right)  
 \label{eq3.74} \\ 
&  - x \ ( p_a p_b) \ g^{\mu \nu} - x \left(1 - 2\beta^2 \right)  p_a^\mu p_b^\nu + x \left( 1- 4 \beta + 2 \beta^2 \right)  p_b^\mu p_a^\nu \Big] \partial_{p_j,\nu} \partial_{p_\gamma,\mu} 
 \nonumber \\
& + \left( p_\gamma \leftrightarrow p_j, \ \beta \leftrightarrow \bar \beta \right) \bigg\} \Bigg ] 
\nonumber \\
    &-  \frac{C_A}{\bar \beta} \Bigg [ \frac{2 \beta f_0(\beta)+4(1-2\beta)}{\bar \beta \beta x}    + \frac{2f_0(\beta) -8\beta}{\bar \beta}   + \frac{f_0(\beta)}{\beta \bar \beta} \Big [g_3(x, \beta) \; p_a^\mu \partial_{p_\gamma, \mu}
     \nonumber \\
    &-g_3(x_1, \bar \beta)\; p_b^\mu \partial_{p_\gamma, \mu}  - g_3(x, \bar \beta)\; p_a^\mu \partial_{p_j, \mu} + g_3(x_1, \beta)\; p_b^\mu \partial_{p_j, \mu}  
     \nonumber \\
    & + (1-x_1)( p_\gamma^\mu \partial_{p_\gamma, \mu} - p_j^\mu \partial_{p_j, \mu}) - 2 p_a^\mu (\bar \beta \partial_{p_\gamma, \mu} + \beta \partial_{p_j, \mu} ) \Big ]\Bigg  ], \nonumber 
\end{align}
with $\bar \beta = 1-\beta$, $x_1 = 1/x$,
\be
\begin{split} 
 &  g_1(x,\beta) = -4 \beta f_0(\beta) +f_1(\beta)  x + \frac{f_2(\beta)}{x},
\\
& g_2(x,\beta) = f_3(1-\beta)  +  \frac{f_3(\beta) }{x^2}, \;\;\; g_3(x, \beta)= (1-\beta)(1-2\beta)(1-x),
\end{split} 
\ee
and 
\be
\begin{split} 
& f_0(\beta) = 1 - 2\beta + 2 \beta^2, \;\;\;
f_1(\beta) = \frac{\beta(14\beta^3 - 30\beta^2 + 19\beta -4)}{1-\beta},
\\
& f_2(\beta) = \frac{-10\beta^4 + 18\beta^3 -5\beta^2 -4\beta +2 }{1-\beta},
\;\;\;
f_3(\beta) = \frac{-2\beta^4 + 3\beta^2 + \beta -1}{\beta(1-\beta)}.
\end{split} 
\ee
Similar to the case of the color-singlet production studied in Ref.~\cite{Agarwal:2025dvo},
the complexity of this result stems from the fact that we keep the observable arbitrary; for any specific observable, the above expression significantly  simplifies.  

\subsection{The case $\yb || b$ }

This case is completely analogous to the $\yb || a $ one described in the previous section. Hence, we discuss  it only very briefly.
The starting point is 
the following expression (c.f. Eq.~(\ref{eq2.17}))
\be
\begin{split} 
  \frac{{\rm d} \sigma^{cb}_{\cal O}}{{\rm d} \tau}  &
  = {\cal N}^{-1} \int  {\rm d} \Phi(p_a,p_b|\tilde p_j,p_\yb, \tilde p_\gamma) |{\cal M}|^2(p_a,p_b;\tilde p_j,p_\yb, \tilde p_\gamma)
    \delta \left ( \tau - \frac{2 p_b p_\yb}{P_b} \right ) {\cal O}(\tilde p_j, \tilde p_\gamma).
\end{split} 
\ee
After performing the boost, the momentum conservation becomes 
\be
p_a + x p_b = p_j + p_\gamma. 
\ee
The jet and photon momenta are parametrized using the Sudakov decomposition as in Eq.~(\ref{eq4.4}) but with $p_b$ replaced with $x p_b$, and $s$ replaced with $x s$.  

Similar to the $\yb || a$ case, the collinear expansion generates power divergences in the $x \to 1$ limit; these divergences are dealt with using integration by parts,  as  discussed in the preceding  section.  Hence, 
without further ado, we just note that  results for the $\yb || b $ case  
can be obtained from 
Eqs~(\ref{eq3.66})
and \eqref{eq3.70} by applying the following replacements 
\be
 p_a \leftrightarrow p_b, \;\;\; \beta \leftrightarrow 1-\beta, \;\;\; P_a \leftrightarrow P_b, \;\;\; {\rm d}\Phi^{xa, b} \leftrightarrow {\rm d}\Phi^{a,xb}.
\ee

\subsection{The case $\yb || \xa$ }
As explained in Section~\ref{sect2}, 
we assume  that the relation between the one-jettiness value $\tau$ and the jet radius $R$ is such,  that when the smallest scalar product is $p_\xa \cdot p_\yb$, partons $\xa$ and $\yb$ are  clustered into a jet.  Hence, in this case, the expression for the cross section reads 
\be
\begin{split} 
  \frac{{\rm d} \sigma_{\cal O}^{\xa \yb}}{{\rm d} \tau}  & = {\cal N}^{-1} \int  {\rm d} \Phi(p_a,p_b|p_\xa,p_\yb,\tilde p_\gamma) |{\cal M}|^2(p_a,p_b;p_\xa,p_\yb, \tilde p_\gamma)
   \theta(p_{\perp,\xa} - p_{\perp,\yb}) \\
&  \times 
   \delta \left ( \tau - \frac{4 p_\xa p_\yb}{P_J} \right ) {\cal O}(p_{[\xa \yb]},\tilde p_\gamma)
    ,
    \label{eq3.93}
\end{split} 
\ee
where $p_{[\xa \yb]} = p_\xa + p_\yb$ is the jet four-momentum.

To simplify Eq.~(\ref{eq3.93}), we use 
the symmetry of the integrand with respect to 
$\xa \leftrightarrow \yb$ 
exchange, to 
remove the transverse momentum ordering $\theta(p_{\perp,\xa} - p_{\perp, \yb})$, and divide the cross section by two.  We then use the momentum mapping described in Appendix~\ref{sec:finalboost} to write Eq.~(\ref{eq3.93}) in the following way
\be
\begin{split} 
\label{eq3.94}
  & \frac{{\rm d} \sigma_{\cal O}^{\xa \yb}}{{\rm d} \tau}   = \frac{{\cal N}^{-1}}{2}
  \frac{{\Omega_\perp^{d-2}}}{4 (2 \pi)^{d-2}}
  \int  {\rm d} \Phi^{ab}_{\gamma j}  
  \; \frac{ {\rm d} s_{\xa \yb}}
  {2 \pi} \;  s_{\xa \yb}^{-\ep} \; \lambda^{1-2\ep}
  \; 
  {\rm d} \alpha_\xa 
\;  [{\rm d} \Omega_\perp^{d-2}] \; 
  ( \alpha_\xa (1- \alpha_\xa) )^{-\ep} 
  \;
  \\
& \times   |{\cal M}|^2(p_a,p_b;p_\xa,p_\yb, \lambda p_\gamma)\;
   \delta \left ( \tau - \frac{2 s_{\xa \yb} }{P_J} \right ) {\cal O}(p_j + (1-\lambda) p_\gamma , \lambda p_\gamma)
   ,
\end{split} 
\ee
where $s_{\xa \yb} = 2 p_\xa \cdot p_\yb$ and  $\lambda = 1-s_{\xa \yb}/s$.
The vectors $p_j$ and $p_\gamma$ are  light-like. 
We emphasize that $p_j$ is \emph{not} the jet momentum as follows from the first argument of the observable function ${\cal O}$.  
Furthermore, the ``original'' photon momentum $\tilde p_\gamma$ 
and the ``final'' photon momentum $p_\gamma$ are proportional, 
but not equal, to each other, i.e. $\tilde p_\gamma = \lambda p_\gamma$.

The rest of the computation involves the expansion of  
the matrix element squared,  and the observable around the collinear limit. To perform it, we use the Sudakov decomposition of $p_{\xa, \yb}$ in terms of $p_{j,\gamma}$.  As shown in  Appendix \ref{sec:finalboost}, 
the following equations hold
\be
\begin{split} 
& p_\xa = \alpha_m p_j + 
\frac{s_{\xa \yb}}{s}(1-\alpha_\xa) p_\gamma  + \sqrt{s_{\xa \yb} \alpha_\xa (1-\alpha_\xa)} n_\perp,\\
& p_\yb = (1-\alpha_\xa) p_j + 
\frac{s_{\xa \yb}}{s} \alpha_\xa  p_\gamma  - \sqrt{s_{\xa \yb} \alpha_\xa (1-\alpha_\xa)} n_\perp,
\end{split}
\label{eq4.17}
\ee
where $n_\perp \cdot p_{j,\gamma} = 0$.

We have to use this decomposition 
in the matrix element 
squared in Eq.~(\ref{eq3.94}) and 
expand it in $s_{\xa \yb} \sim \tau$.
 Such  expansion generates terms of the form 
\be
p_{a,b} \cdot  n_\perp,
\;\;\; (p_{a,b} \cdot n_\perp )^2. 
\ee
We note that integration over directions of $n_\perp$ is possible because 
both the constraint  and the observable do not depend 
on $n_\perp$. Hence, 
\be
p_{x} \cdot  n_\perp 
\to 0, 
\;\;\;\;
(p_{x} \cdot n_\perp )^2
\to -p_{x}^\mu p_{x}^\nu \;
\frac{g_{\perp, \mu \nu}}{2(1-\ep)},
\ee
where $x = a,b$ and 
\be
g_{\perp}^{\mu \nu} 
= g^{\mu \nu} - 
\frac{p_j^\mu {p}_\gamma^\nu + p_j^\nu {p}_\gamma ^\mu }{
p_j \cdot p_\gamma
}.
\ee
Using Eqs~(\ref{eq4.17},\ref{eq4.4}), we easily find 
\be
(p_{x} \cdot n_\perp )^2
\to  \;
\frac{s \beta (1-\beta)}{d-2},\;\;\;\; x = a,b.
\ee

To present the final result for the collinear
$\xa || \yb$ contribution to the cross section, we also need to expand the observable ${\cal O}$. With the required accuracy, we find\footnote{We 
remind the reader that since in this case partons are clustered into a jet, 
one needs to write the observable without assuming $p_j^2=0$, compute the derivative and take the limit $p_j^2 = 0$ only after that. 
}
\be
{\cal O}(p_j + (1-\lambda) p_\gamma , \lambda p_\gamma)
= \left[ 1 
+\frac{s_{\xa \yb}}{s} \; p_{\gamma}^\mu  \left( \partial_{p_j \mu} - \partial_{p_\gamma \mu} \right)
\right ]  {\cal O}(p_j, p_\gamma).
\ee

We are now in a position to 
write the result for the $\xa || \yb$ collinear contribution. While it is straightforward to do so,  there is one peculiar aspect of the outcome of such a calculation that we would like to discuss.

Computing the  expansion of the observable and the matrix element squared, and integrating over $s_{\xa \yb}$, $\alpha_\xa$ and ${\rm d} \Omega_k$ in Eq.~(\ref{eq3.94}),  we find 
leading- and subleading contributions to the cross section in the expansion in $\tau$
\begin{align}
  &    \frac{{\rm d} \sigma^{\rm c_{\xa \yb},{\rm LP}} }{{\rm d} \tau}= \frac{ [\alpha_s] \bar \sigma_0 \; 2^\ep \; P_{J}^{-\ep}
 C_A }{8\tau^{1+\ep}}  
{\rm d} \Phi^{ab}_{\gamma j} \bigg[ - 8  \frac{(1-2 \beta +2\beta^2 )}{ \beta(1-\beta) \ep } 
+\cdots \bigg ]\;  {\cal O}(p_j, p_\gamma),
\\
&     \frac{{\rm d} \sigma^{\rm c_{\xa \yb},{\rm NLP}} }{{\rm d} \tau}= \frac{ [\alpha_s] \bar \sigma_0 \;
  C_A  \; 2^\ep P_{J}^{1-\ep}}{8 s \tau^{\ep}} 
{\rm d} \Phi^{ab}_{\gamma j}
\bigg[   -2     \frac{(4\beta^4 - 8\beta^3 + 2\beta^2 + 2\beta -1 )}{  \beta^2 (1-\beta)^2 \ep}
\nonumber 
\\
& - 4 \frac{(1-2 \beta + 2 \beta^2)}{\beta (1-\beta) \ep }
\; p_{\gamma}^\mu  \left( \partial_{p_j \mu} - \partial_{p_\gamma \mu} \right) + 
 \cdots \bigg ]  {\cal O}(p_j, p_\gamma),
\label{eq4.20}
\end{align}
where ellipses stand for terms without the $\ep \to 0$ poles. 

A peculiar aspect of the above result is that if  we combine the subleading (NLP) contribution in Eq.~(\ref{eq4.20}) with the 
collinear and soft contributions discussed earlier, we do not  immediately observe   
the cancellation of $1/\ep$ poles.  In fact, it only happens  \emph{after} one integrates by parts over $\beta$ in Eq.~(\ref{eq4.20}). 
This integration is particularly simple, because the term without  derivatives of the
observable in Eq.~(\ref{eq4.20}) can be written as 
\be
\frac{(4\beta^4 - 8\beta^3 + 2\beta^2 + 2\beta -1 )}{  \beta^2 (1-\beta)^2}
= \frac{{\rm d} }{{\rm d}{\rm \beta}}
\left (  \frac{ (1-2 \beta + 2 \beta^2)(1-2 \beta) }{\beta (1-\beta)}
\right ).
\ee
Since ${\rm d} \Phi_{\gamma j}^{ab} \sim \beta^{-\ep}(1-\beta)^{-\ep}\; {\rm d} \beta$, integration by parts over $\beta$ is straightforward. We find 
\be
\begin{split}
    \frac{{\rm d} \sigma^{\rm c_{\xa \yb},{\rm NLP}} }{{\rm d} \tau}
    &= \frac{ [\alpha_s] \bar \sigma_0 \;
  C_A  \; 2^\ep P_{J}^{1-\ep}}{4 s \tau^{\ep} \ep} \; 
{\rm d} \Phi^{ab}_{\gamma j} \;
\frac{(1-2  \beta + 2 \beta^2)  }{2 \beta (1-\beta)  }\\
& \times
 \; \bigg\{ -4 p_{\gamma,\mu}  \left( \partial_{p_j}^\mu - \partial_{ p_\gamma}^\mu \right)
+ 2(1-2\beta) \frac{\rm d}{\rm d \beta} 
  \bigg \} \;
  {\cal O}(p_j, p_\gamma) + \cdots \;.
\end{split}
\label{eq3.84}
\ee
Writing the derivative of the observable with respect to $\beta$ as derivatives with respect to $p_{j}$ and $p_\gamma$,  we obtain 
\be
\begin{split}
    & \frac{{\rm d} \sigma^{\rm c_{\xa \yb},{\rm NLP}} }{{\rm d} \tau}
    = \frac{ [\alpha_s] \bar \sigma_0 \;
  C_A  \; 2^\ep P_{J}^{1-\ep}}{4 s \tau^{\ep} \ep} \; 
{\rm d} \Phi^{ab}_{\gamma j} \;
\frac{(1-2  \beta + 2 \beta^2)  }{2 \beta (1-\beta)  }
 \; \bigg\{ -4 p_{\gamma,\mu}  \\
& + (1-2\beta)
      \left[\frac{p_{a, \mu}}{\beta} - \frac{p_{b,\mu}}{(1-\beta)} -  \frac{(1-2\beta)}{\beta(1-\beta)} \; p_{\gamma,\mu} \right] 
  \bigg \}
 \left( \partial_{p_j}^\mu - \partial_{p_\gamma}^\mu \right)
  {\cal O}(p_j, p_\gamma) + \cdots \;.
\end{split}
\label{eq4.35}
\ee
The representation of 
the divergent contribution in Eq.~(\ref{eq4.35}) turns out to be  suitable for establishing the cancellation of the $1/\ep$ poles among all  
next-to-leading-power contributions. 

We quote here the result for the  remaining finite terms in  the $\xa || \yb$ configuration, that we obtain in addition to the divergent ones in Eq.~(\ref{eq4.35})
 \be
 \begin{split}
     & \frac{{\rm d} \sigma^{\rm c_{\xa \yb},{\rm NLP}} }{{\rm d} \tau} \Big |_{\rm fin}
      = \frac{ [\alpha_s] \bar \sigma_0 \;
    P_{J} }{8 s} \; 
 {\rm d} \Phi^{ab}_{\gamma j}  \bigg\{ \frac{C_A}{3} \bigg[\frac{6(1-2\beta+2\beta^2)}{\beta^2 (1-\beta)^2}-\frac{11}{\beta (1-\beta)}-22 \\
 &+ \left(\frac{1}{\beta (1-\beta)} + 22 \right)p_{\gamma,\mu}  \left( \partial_{p_j}^\mu - \partial_{p_\gamma}^\mu \right)\bigg]  -6 C_F \frac{(1- 2\beta+2\beta^2)}{\beta^2 (1-\beta)^2}\bigg\} \; {\cal O}(p_j, p_\gamma).
 \end{split}
 \label{eq3.97}
\ee
We note that we have taken the $\ep \to 0$ limit in the above equation.

\subsection{The final result for the power corrections 
to $q \bar q
\to \gamma + j$
at NLO QCD
}
\label{sec3.6}

Having calculated all the contributions required to obtain  the 
next-to-leading-power corrections 
to the production of a photon and a jet in  the  $q \bar q$ annihilation channel,
we  combine them into the  final result.  The cancellation of the $1/\ep$ poles occurs separately for the clustered and non-clustered cases, leaving the  $\ln \tau$ terms
behind.  The final result is obtained by combining  
\begin{itemize}
\item 
the \emph{clustered} contributions given in   Eqs~(\ref{eq3.41},\ref{eq4.35},\ref{eq3.97});
\item the 
\emph{unclustered} ones from   Eqs~(\ref{eq3.42},\ref{eq3.66},\ref{eq3.70}); 
\item  the 
contribution in Eq.~(\ref{eq3.42a}) that arises because of the modification of the angular distance of  the jet algorithm due to the soft recoil. 
\end{itemize}

We therefore write 
\be
\frac{{\rm d} \sigma_{\cal O}^{\rm NLP}}{{\rm d} \tau}
= \frac{{\rm d} \sigma_{\cal O}^{\rm NLP}}{{\rm d} \tau}\Big |_{\rm cl}
+
\frac{{\rm d} \sigma_{\cal O}^{\rm NLP}}{{\rm d} \tau}\Big |_{\rm nc}
+\frac{{\rm d} \sigma_{\cal O}^{s,R}}{{\rm d} \tau},
\ee
where the last term can be found in Eq.~(\ref{eq3.42a})  and the two other  contributions read 
\be
\begin{split}
 & \frac{ {\rm d } \sigma_{\cal O}^{\rm NLP}}{{\rm d} \tau}
\Big |_{\rm cl}
=  \frac{[\alpha_s] \tilde \sigma_0 P_J}{2 s}
{\rm d} \Phi_{\gamma j}^{ab} \;
\frac{(1-2\beta+2\beta^2)}{2 \beta \bar \beta} \Bigg\{ - \frac{C_F}{\beta \bar \beta} +\frac{C_A}{6} \left( 23 - \frac{22}{1-2\beta+2\beta^2}\right)\\
& - \frac{C_A}{2} \ln \left( \frac{\tau P_J}{2 s} \right) \bigg [ 4 p_\gamma^\mu - (1-2\beta)
      \left(\frac{p_a^\mu}{\beta} - \frac{p_b^\mu}{\bar \beta} -  \frac{(1-2\beta)}{\beta \bar \beta} \; p_\gamma^\mu \right) 
  \bigg ]\left( \partial_{p_j\mu} - \partial_{p_\gamma \mu} \right) \\
& - \frac{C_A}{2} \left[ \frac{11}{3} p_\gamma^\mu - \frac{2(1-2\beta) \beta \bar \beta}{(1-2\beta+2\beta^2)}
      \left(\frac{p_a^\mu}{\beta} - \frac{p_b^\mu}{\bar \beta} -  \frac{(1-2\beta)}{\beta \bar \beta} \; p_\gamma^\mu \right) \right] \left( \partial_{p_j \mu} - \partial_{p_\gamma \mu} \right) \\
& +  \int [{\rm d} \Omega_k] \; {\cal F}_k^{\rm cl} \Bigg\} \; {\cal O}(p_j, p_\gamma),
\end{split}
\label{3.100}
\ee
and
\be
\begin{split}
  & \frac{ {\rm d } \sigma_{\cal O}^{\rm NLP}}{{\rm d} \tau}
\Big |_{\rm nc}
=  \frac{[\alpha_s] \tilde \sigma_0}{s}\;
\frac{(1-2\beta+2\beta^2)}{2 \beta \bar \beta} \Bigg\{P_a \int \limits_0^1  {\rm d} x \; {\rm d} \Phi^{xa, b}_{\gamma j} \; {\cal C}_a(x,\beta, P_a,p_a,p_b,p_j,p_\gamma)\\
& + P_b \int \limits_0^1  {\rm d} x \; {\rm d} \Phi^{a,xb}_{\gamma j} \; {\cal C}_a(x,\bar \beta,P_b,p_b,p_a,p_j,p_\gamma) +
{\rm d} \Phi^{ab}_{\gamma j} \; \int [{\rm d} \Omega_k] \; {\cal F}_k^{\rm nc} \Bigg\} \; {\cal O}(p_j, p_\gamma).
\end{split}
\label{3.101}
\ee

In the above equation, we have introduced the  function ${\cal C}_a$ defined as follows   
\begin{align}
 & {\cal C}_a(x,\beta,P_a,p_a,p_b,p_j,p_\gamma)
= \left[ \delta(1-x) \ln \left( \frac{\tau P_a}{s} \right) -  {\cal L}_0 (1-x) \right] \Bigg \{ \frac{1+\beta}{\bar \beta} C_A \nonumber \\
& + \left( C_F \left( p_j^\mu - \beta p_a^\mu + \bar \beta p_b^\mu \right) + \frac{C_A}{2} \frac{(1+2\beta)p_j^\mu + \beta p_a^\mu - \bar \beta p_b^\mu}{\bar \beta} \right) \partial_{p_j \mu} \nonumber \\
& + \left( C_F \left( p_\gamma^\mu -\bar \beta p_a^\mu + \beta p_b^\mu \right) - \frac{C_A}{2} \frac{(1-2\beta)p_\gamma^\mu - \bar \beta p_a^\mu + \beta p_b^\mu}{\bar \beta} \right) \partial_{p_\gamma \mu}  \Bigg \} \\
&- \delta(1-x)\bigg[ \frac{C_A \beta}{\bar \beta}  \left[ \left(p_j^\mu + \beta p_a^\mu - \bar \beta p_b^\mu \right) \partial_{p_j \mu} + \left( p_\gamma^\mu + \bar \beta p_a^\mu - \beta p_b^\mu \right) \partial_{p_\gamma \mu} \right] \nonumber  \\
 & +2C_F+\frac{2 \beta}{\bar \beta} C_A \Bigg ] + \frac{\beta \bar \beta}{4(1-2\beta+2\beta^2)} \;  R_{\rm ca}(\beta,x, p_a, p_b ) \Bigg\}. \nonumber
\end{align}
Functions ${\cal F}_k^{\rm cl}$, $ {\cal F}_k^{\rm nc}$ and $R_{\rm ca}(\beta,x, p_a, p_b )$ have been already introduced  in Eqs~(\ref{eq3.46},\ref{eq3.57},\ref{eq3.74}), respectively.


\subsection{Numerical checks}
In this section, we provide a numerical validation of the 
next-to-leading-power corrections presented in Section \ref{sec3.6}, focusing 
on partonic cross sections for    various transverse-momenta cuts. 
 Hence, we choose 
\be
{\cal O}(p_j, p_\gamma)= \theta(p_{\perp,j} - p_{\perp,\rm cut}),
\label{3.109}
\ee
and compute the  cross section 
\be
\sigma_{\text{num}}(\tau_{\rm max}, \tau_{\rm min}) = \int \limits_{\tau_{\rm min}}^{\tau_{\rm max} } {\rm d} {\tau} 
\frac{{\rm d} \sigma_{\gamma j}}{{\rm d} \tau} \; {\cal O}(p_j ),
\label{eq3.110}
\ee
for several small values of $\tau_{\rm min}$ and $\tau_{\rm max}$, using the exact  matrix element for $q \bar q \to \gamma + gg$, and the phase space for the three-particle  final state. 
To this end, we implement  the expression shown in Eq.~\eqref{eq1.6}  in a numerical code. Since we work at  
small but non-vanishing  $\tau$, dimensional  regularization is not needed, 
as one-jettiness provides an infra-red cutoff. 
We use $\sqrt{s}= 200$ GeV, $\;P_a=P_b=P_j= \sqrt{s}/2 $, $R=0.4$, and set  $[\alpha_s]$ and $\tilde \sigma_o $ to one.

At the same time, 
 for (sufficiently) small values of $\tau_{\rm min},\; \tau_{\rm max}$,  the same integrated cross section can be computed using leading  and next-to-leading-power contributions derived   in this paper,  
\be
\frac { {\rm d} \sigma_{ \gamma j}
}{{\rm d} \tau}
= \frac { {\rm d} \sigma^{\rm LP}_{\gamma j}
}{{\rm d} \tau}
+ \frac{{\rm d} \sigma_{ \gamma j}^{\rm NLP}}{{\rm d} \tau} + \dots 
\ee
Verifying that the two results actually agree provides a check
on the next-to-leading-power corrections reported in this paper. 

We note that it is   challenging to check the correctness of   the next-to-leading-power corrections with decent accuracy; the reason is that the 
integral in Eq.~(\ref{eq3.110}) 
is dominated by the double- and single-logarithmic, leading-power contributions.  Our strategy 
is to subtract them from 
$\sigma_{\rm num}(\tau_{\rm max},\tau_{\rm min})$ by considering
\be
\bar \sigma_{\rm num}(\tau_{\rm max},\tau_{\rm min}) = \sigma_{\rm num}(\tau_{\rm max},\tau_{\rm min}) - \int \limits_{\tau_{\rm min}}^{\tau_{\rm max} } {\rm d} {\tau} 
\frac{{\rm d} \sigma_{\gamma j}^{\rm LP}}{{\rm d} \tau} \; {\cal O}(p_j ),
\ee
and fit $\bar \sigma$ which  receives contributions from the subleading  terms only. 
To present the results, we write the higher-order power 
corrections in the following form
\be
\begin{split}
   \sqrt{s} \left(  \frac{\mathrm d \sigma_{\gamma j}}{\mathrm d \tau} - \frac{\mathrm d \sigma^{\text{LP}}_{\gamma j}}{\mathrm d \tau} \right) & =  \ln \nu\;C_{\text{NLP,LL}}    + C_{\text{NLP,NLL}} +  
   \nu \ln \nu  \;
   C_{\text{NNLP,LL}}  \\
   & + \nu \;  C_{\text{NNLP,NLL}}   +\cdots,
\end{split}
\label{eq3.109}
\ee
where $\nu = \tau/\sqrt{s}$ and the ellipses indicate the neglected power corrections at higher orders in the 
expansion in $\nu$.  

We  determine the 
$C$-coefficients in  
Eq.~(\ref{eq3.109}) by fitting 
 $\bar \sigma_{\rm num}$  computed for $\nu_{\rm min} = 10^{-5}$ and choosing ${\cal O}(40)$ points for $\nu_{\rm max}$ 
 from the interval 
  $\nu_{\rm max}
 \in [5 \times 10^{-5}, 5 \times 10^{-3}]$.
We note that we do not fit 
all the $C$-coefficients 
in Eq.~(\ref{eq3.109}) 
simultaneously. Instead, we first  extract  the leading-log coefficient $C_{\text{NLP,LL}}$  from data and verify its 
 consistency with the analytic result.   Once this is accomplished, we assume that the $C_{\text{NLP,LL}}$ is correct,  
subtract  it from  $\bar \sigma_{\rm num}(\tau_{\rm max},\tau_{\rm min})$ and fit the difference for the 
coefficient $C_{\rm NLP,NLL}$. 
The value of the obtained coefficient  
$C_{\rm NLP,NLL}$ is  then compared to the analytic results derived this paper, c.f.
Eqs \eqref{3.100} and \eqref{3.101}.  


 \begin{table}[t]
   \centering
    \begin{tabular}{||c|c|c||c|c||}
     \hline \hline
    \multirow{2}{*}{$p_{\perp,\text{cut}}$(GeV)} &\multicolumn{2}{c||}{$C_{\text{NLP,LL}}$} &\multicolumn{2}{c||}{$C_{\text{NLP,NLL}}$} \\ [0.5ex]
    \cline{2-5} 
    & \multicolumn{1}{c|}{analytic} & \multicolumn{1}{c||}{fitted} & \multicolumn{1}{c|}{analytic} & \multicolumn{1}{c||}{fitted} \\
    \hline \hline
     20 & 32.00 & 32.0(3) & 69.27(6) & 71(1)  \\
    \cline{1-5}    
     25 & 20.25  & 20.1(3)  & 31.29(3)  & 31.9(9)   \\
    \cline{1-5} 
     30 & 13.88  & 13.9(3)  &  14.13(3)  & 14.5(8) \\
    \hline
    \end{tabular}
    \caption{Comparison of the subleading coefficients $C_{\text{NLP,LL}}$ and $C_{\text{NLP,NLL}}$ obtained using the fit against the analytic calculation for different cuts on the jet's transverse momenta.
    See text for further details.}
    \label{tab1}
\end{table}

The comparison of the numerical and analytic results  for the 
power corrections   is shown  in Table \ref{tab1} for different values of the transverse-momentum  cut.
It follows from Table \ref{tab1} that  the agreement is quite impressive, especially given the smallness of the sub-leading contributions in 
the region of the fit. 

\begin{figure}[h!]
    \centering
    \includegraphics[scale=0.7]{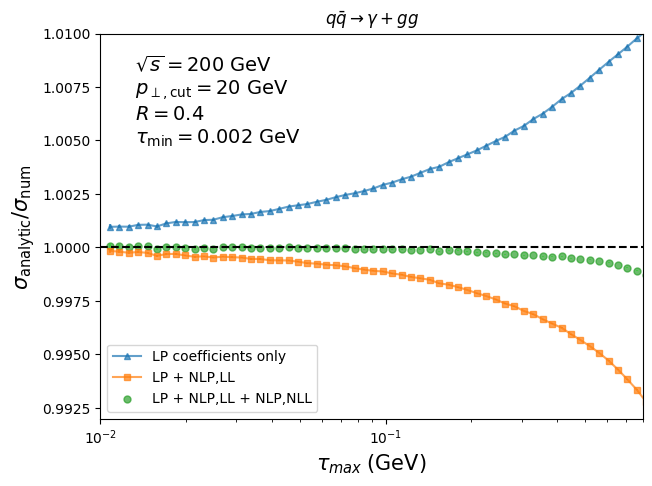}
    \caption{
    The comparison of the ``exact'' cross section 
    $\sigma_{\rm num}$ (c.f. Eq.~(\ref{eq3.110}) and 
    its  various approximations 
    obtained by different truncations of  the 
    expansion in small $N$-jettiness.
The analytic approximations including the leading-power (LP) contributions, the LP + leading-logarithmic (LL) next-to-leading-power (NLP) correction, and the LP + full NLP corrections.  The three curves in the plot become indistinguishable   for $\tau < 10^{-3}$.}
    \label{fig1}
\end{figure}

Another useful illustration of the correctness of the next-to-leading-power corrections computed in this paper is provided 
in Figure \ref{fig1}. There we plot ratios of   analytic and numerical NLO cross sections 
$\sigma_{\rm num}(\tau_{\rm max},\tau_{\rm min})$; the important point is that 
different number of terms in the $\tau$-expansion are retained in  the   analytic results shown there. 
It is clear  from the plot that the inclusion of full next-to-leading-power corrections 
extends the region of the  
$\nu$ values, where  the numerical 
and analytical results agree, indicating the correctness 
of the latter. 


\section{Conclusions}
\label{sect:conclusions}

In this paper we have derived, for the very first time, the subleading power corrections in the one-jettiness variable to a process with the final-state jet. We focused on the partonic process $q \bar q \to \gamma + j$ since it is sufficiently simple  to directly work with the relevant  matrix elements, and it does not require  the photon-isolation procedure to get a physical result.   We employed a fully-realistic $k_\perp$ jet algorithm in this study. 

We have shown that the method for computing power corrections  developed  by us in Ref.~\cite{Agarwal:2025dvo} to describe production 
of arbitrary color-singlet final states in hadron collisions,  remains effective also for processes with final-state jets.  Key elements of this approach are  momenta redefinitions and Lorentz transformations; they are familiar from the discussion  of general subtraction schemes 
at NLO and NNLO (see  \cite{DelDuca:2019ctm} and references therein). 

Our study of power corrections  can be extended in several ways in the future.  
First, in this paper we have relied on the explicit form of the matrix element and did not attempt to design a process-independent  framework similar to what has been done in Ref.~\cite{Agarwal:2025dvo} for the color-singlet final states. It will be interesting to understand  how to generalize this approach to final states with arbitrary number of jets, where the analytic  expressions  for relevant  matrix elements cannot be used.

Second, it is worthwhile to extend the current analysis to processes with an on-shell vector boson in the final state.  Although  such an extension should be  straightforward, the gauge-boson on-shell  constraint may require some care with Lorentz transformations and momenta redefinitions.

Third, the major reason for the complicated   analytic expressions for the  power corrections is  the derivatives of  observables. For this reason,  it will  be useful to design a framework that will allow one  to treat them as changes in kinematics of observable quantities in a more universal and easy-to-handle way. 

Finally, it would be interesting to extend the analysis of power corrections in the $N$-jettiness variable to 
next-to-next-to-leading order. Although the complexity of this task remains outstanding, we hope that the improved understanding of 
the power corrections provided by this paper and also by Ref.~\cite{Agarwal:2025dvo} constitutes a good  starting point for 
attempting it. 

\acknowledgments
This research was  supported by the Deutsche Forschungsgemeinschaft (DFG, German Research Foundation) under grant no.\ 396021762 - TRR 257. 

\appendix

\section{Phase-space  parametrization
for the final-state collinear limit
}
\label{sec:finalboost}
In this appendix, we derive  the momenta mapping and the phase-space parametrization that is suitable for describing the  final-state collinear limit.  The goal is to map the momentum
conservation condition
\be
p_a+p_b = p_\xa + p_\yb + \tilde p_\gamma, 
\ee
onto
\be
p_a + p_b = p_j + p_\gamma, 
\ee
where $p_j^2 = 0$ and $p_\gamma^2 = \tilde p_\gamma^2 =0$.  The momentum $p_j$ is  related to the 
momentum of the final-state jet, but it is not identical to it. 

To construct the 
momentum $p_j$,  we write it as a linear combination of
two vectors $p_{\xa \yb} = p_\xa + p_\yb$ and $P_{ab} =p_a+p_b$, 
\be
p_j = \frac{1}{\lambda}  \left ( p_{\xa \yb} - \frac{ p_{\xa \yb} \cdot P_{a b}}{P_{ab}^2} P_{ab}  \right )
 +  x P_{a b}.
\label{eq2.3}
 \ee 
Since the four-vector in brackets is orthogonal to $P_{ab}$, we find 
\be
x = \frac{p_j \cdot P_{ab}}{P_{ab}^2} 
 = \frac{1}{2}.
\ee
Furthermore, using 
$p_\gamma^2 = \tilde p_\gamma^2=0$, we obtain 
\be
P_{ab} \cdot p_j = P_{ab} \cdot p_{\xa \yb} - \frac{1}{2} s_{\xa \yb},
\ee
where $s_{\xa \yb} = 2 p_{\xa} \cdot p_{ \yb}$.

The parameter $\lambda$ in Eq.~(\ref{eq2.3})  is adjusted to ensure that $p_j^2 = 0$. We find
\be
\lambda = 1 - \frac{s_{\xa \yb}}{P_{ab}^2}.
\label{eq2.4}
\ee
Finally, using Eq.~(\ref{eq2.3}), 
we express  $p_{\xa \yb}$ in terms of $p_j$
\be
p_{\xa \yb} 
= \lambda p_j + (1-\lambda) P_{ab},
\label{eq2.8}
\ee
which immediately implies the following relation for the photon momenta 
\be
\tilde p_\gamma = \lambda p_\gamma.
\label{eqa.9}
\ee

Our goal is to rewrite   
the phase space for $\xa, \yb$ and $\tilde \gamma$ 
in such a way that expansion in $s_{\xa \yb} \sim \tau$ at fixed $p_{j,\gamma}$ becomes possible. 
Below we sketch the derivation of the relevant formula; 
its detailed discussion in a broader context can be found in 
Ref.~\cite{Catani:2000ef}.

We begin by  writing 
\be
\begin{split} 
& \int [{\rm d} p_\xa] [{\rm d} p_\yb] [{\rm d} \tilde p_\gamma] (2\pi)^d \delta(P_{ab} - p_\xa - p_\yb - \tilde p_\gamma)
 = 
\\
& 
\int \frac{{\rm d} s_{\xa \yb}}{2 \pi} \;
 [{\rm d} p_{\xa \yb}][{\rm d} \tilde p_\gamma] (2\pi)^d \delta(P_{ab} - p_{\xa \yb} - \tilde p_\gamma)
\;   \int [{\rm d} p_\xa] [{\rm d} p_\yb] (2 \pi)^d \delta(p_{\xa \yb} - p_\xa - p_\yb),
\end{split} 
     \ee
As the next step, we consider the integral 
over $[{\rm d} p_{\xa \yb}][{\rm d} \tilde p_\gamma]$
in the rest frame of $P_{ab}$ and find  
\be
[{\rm d} p_{\xa \yb}][{\rm d} \tilde p_\gamma] (2\pi)^d \delta(P_{ab} - p_{\xa \yb} - \tilde p_\gamma)
= {\rm d} \Omega_{\tilde \gamma} \;
{\cal N} \left ( 1 - \frac{s_{\xa \yb}}{P_{ab}^2} \right )^{d-3},
\ee
where ${\cal N}$ is a function of 
$P_{ab}^2$ only, and ${\rm d}\Omega_{\tilde \gamma}$ is the solid angle that parametrizes the direction of the  photon momentum 
$\vec p_{\tilde \gamma}$ 
or, equivalently, of $\vec p_{\xa \yb}$.
To relate this result to 
the phase-space of $p_j$ and $p_\gamma$,
we use  Eq.~(\ref{eq2.3}).
It follows from that  equation that in the rest frame of $P_{ab}$,  the directions of $\vec p_j$ and 
$\vec p_{\xa \yb}$  coincide. 
Thus, 
\be
[{\rm d} p_{\xa \yb}][{\rm d} \tilde p_\gamma] (2\pi)^d \delta(P_{ab} - p_{\xa \yb} - \tilde p_\gamma)
=\lambda^{d-3} 
[{\rm d} p_{j}][{\rm d} p_\gamma] (2\pi)^d \delta(P_{ab} - p_j - p_\gamma).
\ee
The  relation between $p_\gamma$ and $\tilde p_\gamma$ is  given in 
Eq.~(\ref{eqa.9}).  Putting everything together, 
we arrive at the final formula for the 
phase space that is  suitable for describing the collinear limit
\be
\begin{split} 
\int [{\rm d} p_\xa] [{\rm d} p_\yb] [{\rm d} \tilde p_\gamma] (2\pi)^d \delta(P_{ab} & - p_\xa - p_\yb  - \tilde p_\gamma)
 =  \int [{\rm d} p_j][{\rm d} p_\gamma] (2\pi)^d \delta(P_{ab} - p_j - p_\gamma)
 \\
& \times \int \limits_{0}^{s} \;
\frac{ {\rm d} s_{\xa \yb}}{2 \pi} \; \lambda^{d-3} \; 
    \int [{\rm d} p_\xa] [{\rm d} p_\yb] (2 \pi)^d \delta(p_{\xa \yb} - p_\xa - p_\yb).
  \end{split} 
 \label{eq3.10}
     \ee

To use  this formula for computing power corrections, we need to understand how 
to integrate over 
$p_\xa$ and $p_\yb$ in the vicinity of the collinear $\xa || \yb$ limit. To this end, we use 
$p_j$ and $p_\gamma$ as basis 
vectors for  the Sudakov decomposition of $p_\xa$ and $p_\yb$. We find 
     \be
     \begin{split}
       & p_{\xa} = \alpha_{\xa}  p_j + \beta_{\xa}  p_\gamma + p_\perp,
       \\
       & p_{\yb} = \alpha_{\yb}  p_j + \beta_{\yb}  p_\gamma - p_\perp.
       \end{split}
       \label{eqa.11}
    \ee
  If the invariant mass $s_{\xa \yb} = 2 p_\xa \cdot p_\yb$ is small,  and $p_j$ is the collinear direction, then $\alpha_\xa  \sim \alpha_n \sim 1$, $\beta_{\xa,\yb} \sim s_{\xa \yb}/P_{ab}^2$ and
    $|p_\perp| \sim \sqrt{s_{\xa \yb}/P_{ab}^2}$. 
Using the Sudakov decomposition, 
we easily find the following parametrization of the $(\xa \yb)$ phase space 
\be
\begin{split}
& \int [{\rm d} p_\xa] [{\rm d} p_\yb] (2 \pi)^d \delta(p_{\xa \yb} - p_\xa - p_\yb)
\\
& 
=  
 \frac{s_{\xa \yb}^{-\ep} \; \Omega_\perp^{(d-2)} }{4 (2\pi)^{d-2}}
      \int \limits_{0}^{1} {\rm d} \alpha_\xa \; 
      [{\rm d} \Omega_\perp^{(d-2)}]
      \left ( \alpha_\xa  ( 
      1- \alpha_{\xa} )
      \right )^{-\ep},
\end{split}
\ee
where the azimuthal angle describes directions of the vector 
$p_\perp$ in Eq.~(\ref{eqa.11}).
With this parametrization, it is possible to expand the explicit matrix element squared for $q \bar q \to \gamma gg$ in the collinear $\xa || \yb$ kinematics.  Indeed,  since 
\be
\begin{split} 
& p_\xa = \alpha_m p_j + 
\frac{s_{\xa \yb}}{s}(1-\alpha_m) p_\gamma + \sqrt{s_{\xa \yb} \alpha_m (1-\alpha_m)} n_\perp,\\
& p_\yb = (1-\alpha_m) p_j + 
\frac{s_{\xa \yb}}{s} \alpha_m  p_\gamma - \sqrt{s_{\xa \yb} \alpha_m (1-\alpha_m)} n_\perp,
\end{split}
\ee
and $s_{\xa \yb} \sim \tau$,  it is straightforward to construct the expansion through next-to-leading power.

\section{A shift in 
$R_{\xa \yb}$}
\label{app:shift}

In this appendix, we discuss    
the change in $R_{\xa \yb}$ induced by  the soft boost.   According to our notation, the momentum of the parton $\xa$ is a  boosted and rescaled $p_j$, 
whereas the parton $\yb$ is assigned the momentum $k$. 
Then, the following relation between $p_\xa$,  $p_j$ and $k$ holds 
\be
p^\mu_\xa = \left (1 - \frac{\omega_k}{2 E_j} 
\right )p^\mu_j
-\frac{k^\mu}{2} + \frac{k \cdot p_j}{2 E_j} t^\mu,
\ee
where $t^\mu = (1, \vec 0)$, we  work in the center-of-mass frame of colliding partons $a$ and $b$, $E_j$ is the energy of $j,\gamma,a,b$,  and $\omega_k$ is the energy of the parton $\xa$.
We can rewrite the above formula in the following way
\be
p_\xa
= \left (1 - \frac{\omega_k}{2E_j} (1 + \cos \theta_{kj} ) 
\right ) p_j 
- \frac{\omega_k}{2}  \left ( 0, 
\vec n_k - \cos \theta_{kj} \vec n_j 
\right ).
\ee
We can also write  $p_\xa$ as 
follows 
\be
p_\xa = \frac{E_\xa}{E_j} p_j + E_j \left ( 0, \vec n_\xa - \vec n_j \right ), 
\ee
where we work to first order in the difference between $p_\xa$ and $p_j$ caused by the emission of a gluon. 

We can match the two equations if we choose 
\be
E_\xa = E_j 
\left (1 - \frac{\omega_k}{2E_j} (1 + \cos \theta_{kj} ) 
\right ),
\ee
and 
\be
\vec n_\xa = \vec n_j 
-\frac{\omega_k}{2 E_j} 
\left ( \vec n_k - \cos \theta_{kj} \vec n_j \right ).
\ee

We assume that vectors $\vec n_\xa$ and $\vec n_j$ are parametrized as follows 
\begin{equation}
\begin{split}
 \vec n_x = 
(\sin \theta_x \cos \varphi_x, \sin \theta_x \sin \varphi_x, \cos \theta_x),
\end{split}
\end{equation}
where $x= \xa,j$, and the $z$-axis is aligned with the vector $\vec n_a$.
Then, 
\be
[\vec n_\xa \times \vec n_j] \cdot \vec n_a =
\sin \theta_m \sin \theta_j \sin \left (\varphi_j - \varphi_m \right).
\ee
At the same time, 
\be
[\vec n_\xa \times \vec n_j] \cdot \vec n_a = 
-\frac{\omega_k}{2 E_j}
[ \vec n_k \times \vec n_j] \cdot \vec n_a.
\ee
Since $\theta_m \sim \theta_j$, we easily find 
\be
\varphi_\xa - \varphi_j
\approx \frac{\omega_k}{2 E_j \sin^2 \theta_j} 
\; [ \vec n_k \times \vec n_j] \cdot \vec n_a 
+{\cal O}(\omega_k^2).
\ee
Similarly, 
\be
\theta_m -  \theta_j 
\approx \frac{\omega_k}{2E_j \sin \theta_j} 
[ \vec n_k \times \vec n_j]
\cdot 
[\vec n_a \times \vec n_j].
\ee

We can use these results to derive the difference 
between $R_{\xa \yb}$ and 
$R_{j \gamma}$. Expanding in Taylor series, we obtain 
\be
R_{\xa \yb} 
= R_{j k}
+ \frac{\omega_k}{2 E_j \sin^2 \theta_j}
[\vec n_k \times \vec n_j]
\cdot 
\left ( 
\frac{\partial R_{jk} }{\partial \varphi_j} \vec n_a 
- \frac{\partial R_{jk}}{\partial \eta_j} [ \vec n_a \times \vec n_j] 
\right )
+{\cal O}(\omega_k^2).
\ee
For the jet algorithm in Eq.~(\ref{eq1.1}), we find 
\be
- \frac{\partial R_{jk}}{\partial \varphi_j}
=\frac{ f_\varphi(\varphi_{jk})  \;{\rm sgn}(\sin \varphi_{jk})}{R_{jk}},
\ee
where $\varphi_{jk} = \varphi_j - \varphi_k$.

\vspace{10mm}

\bibliographystyle{JHEP}
\bibliography{rref}

\providecommand{\href}[2]{#2}\begingroup\raggedright\begin{thebibliography}{100}

\bibitem{Gehrmann:2015bfy}
T.~Gehrmann, J.M.~Henn and N.A.~Lo~Presti, \emph{{Analytic form of the two-loop planar five-gluon all-plus-helicity amplitude in QCD}}, \href{https://doi.org/10.1103/PhysRevLett.116.062001}{\emph{Phys. Rev. Lett.} {\bfseries 116} (2016) 062001} [\href{https://arxiv.org/abs/1511.05409}{{\ttfamily 1511.05409}}].

\bibitem{Chicherin:2018mue}
D.~Chicherin, T.~Gehrmann, J.M.~Henn, N.A.~Lo~Presti, V.~Mitev and P.~Wasser, \emph{{Analytic result for the nonplanar hexa-box integrals}}, \href{https://doi.org/10.1007/JHEP03(2019)042}{\emph{JHEP} {\bfseries 03} (2019) 042} [\href{https://arxiv.org/abs/1809.06240}{{\ttfamily 1809.06240}}].

\bibitem{Abreu:2018rcw}
S.~Abreu, B.~Page and M.~Zeng, \emph{{Differential equations from unitarity cuts: nonplanar hexa-box integrals}}, \href{https://doi.org/10.1007/JHEP01(2019)006}{\emph{JHEP} {\bfseries 01} (2019) 006} [\href{https://arxiv.org/abs/1807.11522}{{\ttfamily 1807.11522}}].

\bibitem{Abreu:2021oya}
S.~Abreu, F.~Febres~Cordero, H.~Ita, B.~Page and V.~Sotnikov, \emph{{Leading-color two-loop QCD corrections for three-jet production at hadron colliders}}, \href{https://doi.org/10.1007/JHEP07(2021)095}{\emph{JHEP} {\bfseries 07} (2021) 095} [\href{https://arxiv.org/abs/2102.13609}{{\ttfamily 2102.13609}}].

\bibitem{Chicherin:2018old}
D.~Chicherin, T.~Gehrmann, J.M.~Henn, P.~Wasser, Y.~Zhang and S.~Zoia, \emph{{All Master Integrals for Three-Jet Production at Next-to-Next-to-Leading Order}}, \href{https://doi.org/10.1103/PhysRevLett.123.041603}{\emph{Phys. Rev. Lett.} {\bfseries 123} (2019) 041603} [\href{https://arxiv.org/abs/1812.11160}{{\ttfamily 1812.11160}}].

\bibitem{Abreu:2020jxa}
S.~Abreu, H.~Ita, F.~Moriello, B.~Page, W.~Tschernow and M.~Zeng, \emph{{Two-Loop Integrals for Planar Five-Point One-Mass Processes}}, \href{https://doi.org/10.1007/JHEP11(2020)117}{\emph{JHEP} {\bfseries 11} (2020) 117} [\href{https://arxiv.org/abs/2005.04195}{{\ttfamily 2005.04195}}].

\bibitem{Canko:2020ylt}
D.D.~Canko, C.G.~Papadopoulos and N.~Syrrakos, \emph{{Analytic representation of all planar two-loop five-point Master Integrals with one off-shell leg}}, \href{https://doi.org/10.1007/JHEP01(2021)199}{\emph{JHEP} {\bfseries 01} (2021) 199} [\href{https://arxiv.org/abs/2009.13917}{{\ttfamily 2009.13917}}].

\bibitem{Abreu:2021smk}
S.~Abreu, H.~Ita, B.~Page and W.~Tschernow, \emph{{Two-loop hexa-box integrals for non-planar five-point one-mass processes}}, \href{https://doi.org/10.1007/JHEP03(2022)182}{\emph{JHEP} {\bfseries 03} (2022) 182} [\href{https://arxiv.org/abs/2107.14180}{{\ttfamily 2107.14180}}].

\bibitem{Kardos:2022tpo}
A.~Kardos, C.G.~Papadopoulos, A.V.~Smirnov, N.~Syrrakos and C.~Wever, \emph{{Two-loop non-planar hexa-box integrals with one massive leg}}, \href{https://doi.org/10.1007/JHEP05(2022)033}{\emph{JHEP} {\bfseries 05} (2022) 033} [\href{https://arxiv.org/abs/2201.07509}{{\ttfamily 2201.07509}}].

\bibitem{Abreu:2023rco}
S.~Abreu, D.~Chicherin, H.~Ita, B.~Page, V.~Sotnikov, W.~Tschernow et~al., \emph{{All Two-Loop Feynman Integrals for Five-Point One-Mass Scattering}}, \href{https://doi.org/10.1103/PhysRevLett.132.141601}{\emph{Phys. Rev. Lett.} {\bfseries 132} (2024) 141601} [\href{https://arxiv.org/abs/2306.15431}{{\ttfamily 2306.15431}}].

\bibitem{Chicherin:2021dyp}
D.~Chicherin, V.~Sotnikov and S.~Zoia, \emph{{Pentagon functions for one-mass planar scattering amplitudes}}, \href{https://doi.org/10.1007/JHEP01(2022)096}{\emph{JHEP} {\bfseries 01} (2022) 096} [\href{https://arxiv.org/abs/2110.10111}{{\ttfamily 2110.10111}}].

\bibitem{DeLaurentis:2023izi}
G.~De~Laurentis, H.~Ita and V.~Sotnikov, \emph{{Double-virtual NNLO QCD corrections for five-parton scattering. II. The quark channels}}, \href{https://doi.org/10.1103/PhysRevD.109.094024}{\emph{Phys. Rev. D} {\bfseries 109} (2024) 094024} [\href{https://arxiv.org/abs/2311.18752}{{\ttfamily 2311.18752}}].

\bibitem{DeLaurentis:2023nss}
G.~De~Laurentis, H.~Ita, M.~Klinkert and V.~Sotnikov, \emph{{Double-virtual NNLO QCD corrections for five-parton scattering. I. The gluon channel}}, \href{https://doi.org/10.1103/PhysRevD.109.094023}{\emph{Phys. Rev. D} {\bfseries 109} (2024) 094023} [\href{https://arxiv.org/abs/2311.10086}{{\ttfamily 2311.10086}}].

\bibitem{Agarwal:2024jyq}
B.~Agarwal, G.~Heinrich, S.P.~Jones, M.~Kerner, S.Y.~Klein, J.~Lang et~al., \emph{{Two-loop amplitudes for $ t\overline{t}H $ production: the quark-initiated N$_{f}$-part}}, \href{https://doi.org/10.1007/JHEP05(2024)013}{\emph{JHEP} {\bfseries 05} (2024) 013} [\href{https://arxiv.org/abs/2402.03301}{{\ttfamily 2402.03301}}].

\bibitem{Badger:2024dxo}
S.~Badger, M.~Becchetti, C.~Brancaccio, H.B.~Hartanto and S.~Zoia, \emph{{Numerical evaluation of two-loop QCD helicity amplitudes for $ gg\to t\overline{t}g $ at leading colour}}, \href{https://doi.org/10.1007/JHEP03(2025)070}{\emph{JHEP} {\bfseries 03} (2025) 070} [\href{https://arxiv.org/abs/2412.13876}{{\ttfamily 2412.13876}}].

\bibitem{Agarwal:2023suw}
B.~Agarwal, F.~Buccioni, F.~Devoto, G.~Gambuti, A.~von Manteuffel and L.~Tancredi, \emph{{Five-parton scattering in QCD at two loops}}, \href{https://doi.org/10.1103/PhysRevD.109.094025}{\emph{Phys. Rev. D} {\bfseries 109} (2024) 094025} [\href{https://arxiv.org/abs/2311.09870}{{\ttfamily 2311.09870}}].

\bibitem{FebresCordero:2023pww}
F.~Febres~Cordero, G.~Figueiredo, M.~Kraus, B.~Page and L.~Reina, \emph{{Two-loop master integrals for leading-color $ pp\to t\overline{t}H $ amplitudes with a light-quark loop}}, \href{https://doi.org/10.1007/JHEP07(2024)084}{\emph{JHEP} {\bfseries 07} (2024) 084} [\href{https://arxiv.org/abs/2312.08131}{{\ttfamily 2312.08131}}].

\bibitem{DeLaurentis:2025dxw}
G.~De~Laurentis, H.~Ita, B.~Page and V.~Sotnikov, \emph{{Compact two-loop QCD corrections for Vjj production in proton collisions}}, \href{https://doi.org/10.1007/JHEP06(2025)093}{\emph{JHEP} {\bfseries 06} (2025) 093} [\href{https://arxiv.org/abs/2503.10595}{{\ttfamily 2503.10595}}].

\bibitem{Gehrmann:2018yef}
T.~Gehrmann, J.M.~Henn and N.A.~Lo~Presti, \emph{{Pentagon functions for massless planar scattering amplitudes}}, \href{https://doi.org/10.1007/JHEP10(2018)103}{\emph{JHEP} {\bfseries 10} (2018) 103} [\href{https://arxiv.org/abs/1807.09812}{{\ttfamily 1807.09812}}].

\bibitem{Bargiela:2021wuy}
P.~Bargiela, F.~Caola, A.~von Manteuffel and L.~Tancredi, \emph{{Three-loop helicity amplitudes for diphoton production in gluon fusion}}, \href{https://doi.org/10.1007/JHEP02(2022)153}{\emph{JHEP} {\bfseries 02} (2022) 153} [\href{https://arxiv.org/abs/2111.13595}{{\ttfamily 2111.13595}}].

\bibitem{Caola:2021rqz}
F.~Caola, A.~Chakraborty, G.~Gambuti, A.~von Manteuffel and L.~Tancredi, \emph{{Three-loop helicity amplitudes for four-quark scattering in massless QCD}}, \href{https://doi.org/10.1007/JHEP10(2021)206}{\emph{JHEP} {\bfseries 10} (2021) 206} [\href{https://arxiv.org/abs/2108.00055}{{\ttfamily 2108.00055}}].

\bibitem{Caola:2021izf}
F.~Caola, A.~Chakraborty, G.~Gambuti, A.~von Manteuffel and L.~Tancredi, \emph{{Three-Loop Gluon Scattering in QCD and the Gluon Regge Trajectory}}, \href{https://doi.org/10.1103/PhysRevLett.128.212001}{\emph{Phys. Rev. Lett.} {\bfseries 128} (2022) 212001} [\href{https://arxiv.org/abs/2112.11097}{{\ttfamily 2112.11097}}].

\bibitem{Canko:2021xmn}
D.D.~Canko and N.~Syrrakos, \emph{{Planar three-loop master integrals for 2 {\textrightarrow} 2 processes with one external massive particle}}, \href{https://doi.org/10.1007/JHEP04(2022)134}{\emph{JHEP} {\bfseries 04} (2022) 134} [\href{https://arxiv.org/abs/2112.14275}{{\ttfamily 2112.14275}}].

\bibitem{Gehrmann:2023jyv}
T.~Gehrmann, P.~Jakub{\v{c}}{\'\i}k, C.C.~Mella, N.~Syrrakos and L.~Tancredi, \emph{{Planar three-loop QCD helicity amplitudes for V+jet production at hadron colliders}}, \href{https://doi.org/10.1016/j.physletb.2023.138369}{\emph{Phys. Lett. B} {\bfseries 848} (2024) 138369} [\href{https://arxiv.org/abs/2307.15405}{{\ttfamily 2307.15405}}].

\bibitem{Chen:2025utl}
X.~Chen, X.~Guan and B.~Mistlberger, \emph{{Three-Loop QCD corrections to the production of a Higgs boson and a Jet}},  \href{https://arxiv.org/abs/2504.06490}{{\ttfamily 2504.06490}}.

\bibitem{Gehrmann:2024tds}
T.~Gehrmann, J.~Henn, P.~Jakub{\v{c}}{\'\i}k, J.~Lim, C.C.~Mella, N.~Syrrakos et~al., \emph{{Graded transcendental functions: an application to four-point amplitudes with one off-shell leg}}, \href{https://doi.org/10.1007/JHEP12(2024)215}{\emph{JHEP} {\bfseries 12} (2024) 215} [\href{https://arxiv.org/abs/2410.19088}{{\ttfamily 2410.19088}}].

\bibitem{Liu:2024ont}
Y.~Liu, A.~Matija{\v{s}}i{\'c}, J.~Miczajka, Y.~Xu, Y.~Xu and Y.~Zhang, \emph{{Analytic computation of three-loop five-point Feynman integrals}}, \href{https://doi.org/10.1103/qrk2-cym5}{\emph{Phys. Rev. D} {\bfseries 112} (2025) 016021} [\href{https://arxiv.org/abs/2411.18697}{{\ttfamily 2411.18697}}].

\bibitem{Henn:2023vbd}
J.M.~Henn, J.~Lim and W.J.~Torres~Bobadilla, \emph{{First look at the evaluation of three-loop non-planar Feynman diagrams for Higgs plus jet production}}, \href{https://doi.org/10.1007/JHEP05(2023)026}{\emph{JHEP} {\bfseries 05} (2023) 026} [\href{https://arxiv.org/abs/2302.12776}{{\ttfamily 2302.12776}}].

\bibitem{DiVita:2014pza}
S.~Di~Vita, P.~Mastrolia, U.~Schubert and V.~Yundin, \emph{{Three-loop master integrals for ladder-box diagrams with one massive leg}}, \href{https://doi.org/10.1007/JHEP09(2014)148}{\emph{JHEP} {\bfseries 09} (2014) 148} [\href{https://arxiv.org/abs/1408.3107}{{\ttfamily 1408.3107}}].

\bibitem{Long:2024bmi}
M.-M.~Long, \emph{{Three-loop ladder diagrams with two off-shell legs}}, \href{https://doi.org/10.1007/JHEP01(2025)018}{\emph{JHEP} {\bfseries 01} (2025) 018} [\href{https://arxiv.org/abs/2410.15431}{{\ttfamily 2410.15431}}].

\bibitem{Gehrmann-DeRidder:2005btv}
A.~Gehrmann-De~Ridder, T.~Gehrmann and E.W.N.~Glover, \emph{{Antenna subtraction at NNLO}}, \href{https://doi.org/10.1088/1126-6708/2005/09/056}{\emph{JHEP} {\bfseries 09} (2005) 056} [\href{https://arxiv.org/abs/hep-ph/0505111}{{\ttfamily hep-ph/0505111}}].

\bibitem{Caola:2017dug}
F.~Caola, K.~Melnikov and R.~R\"ontsch, \emph{{Nested soft-collinear subtractions in NNLO QCD computations}}, \href{https://doi.org/10.1140/epjc/s10052-017-4774-0}{\emph{Eur. Phys. J. C} {\bfseries 77} (2017) 248} [\href{https://arxiv.org/abs/1702.01352}{{\ttfamily 1702.01352}}].

\bibitem{Currie:2013vh}
J.~Currie, E.W.N.~Glover and S.~Wells, \emph{{Infrared Structure at NNLO Using Antenna Subtraction}}, \href{https://doi.org/10.1007/JHEP04(2013)066}{\emph{JHEP} {\bfseries 04} (2013) 066} [\href{https://arxiv.org/abs/1301.4693}{{\ttfamily 1301.4693}}].

\bibitem{DelDuca:2016csb}
V.~Del~Duca, C.~Duhr, A.~Kardos, G.~Somogyi and Z.~Tr\'ocs\'anyi, \emph{{Three-Jet Production in Electron-Positron Collisions at Next-to-Next-to-Leading Order Accuracy}}, \href{https://doi.org/10.1103/PhysRevLett.117.152004}{\emph{Phys. Rev. Lett.} {\bfseries 117} (2016) 152004} [\href{https://arxiv.org/abs/1603.08927}{{\ttfamily 1603.08927}}].

\bibitem{DelDuca:2016ily}
V.~Del~Duca, C.~Duhr, A.~Kardos, G.~Somogyi, Z.~Sz\H{o}r, Z.~Tr\'ocs\'anyi et~al., \emph{{Jet production in the CoLoRFulNNLO method: event shapes in electron-positron collisions}}, \href{https://doi.org/10.1103/PhysRevD.94.074019}{\emph{Phys. Rev. D} {\bfseries 94} (2016) 074019} [\href{https://arxiv.org/abs/1606.03453}{{\ttfamily 1606.03453}}].

\bibitem{Czakon:2010td}
M.~Czakon, \emph{{A novel subtraction scheme for double-real radiation at NNLO}}, \href{https://doi.org/10.1016/j.physletb.2010.08.036}{\emph{Phys. Lett. B} {\bfseries 693} (2010) 259} [\href{https://arxiv.org/abs/1005.0274}{{\ttfamily 1005.0274}}].

\bibitem{Czakon:2011ve}
M.~Czakon, \emph{{Double-real radiation in hadronic top quark pair production as a proof of a certain concept}}, \href{https://doi.org/10.1016/j.nuclphysb.2011.03.020}{\emph{Nucl. Phys. B} {\bfseries 849} (2011) 250} [\href{https://arxiv.org/abs/1101.0642}{{\ttfamily 1101.0642}}].

\bibitem{Czakon:2014oma}
M.~Czakon and D.~Heymes, \emph{{Four-dimensional formulation of the sector-improved residue subtraction scheme}}, \href{https://doi.org/10.1016/j.nuclphysb.2014.11.006}{\emph{Nucl. Phys. B} {\bfseries 890} (2014) 152} [\href{https://arxiv.org/abs/1408.2500}{{\ttfamily 1408.2500}}].

\bibitem{Catani:2007vq}
S.~Catani and M.~Grazzini, \emph{{An NNLO subtraction formalism in hadron collisions and its application to Higgs boson production at the LHC}}, \href{https://doi.org/10.1103/PhysRevLett.98.222002}{\emph{Phys. Rev. Lett.} {\bfseries 98} (2007) 222002} [\href{https://arxiv.org/abs/hep-ph/0703012}{{\ttfamily hep-ph/0703012}}].

\bibitem{Jouttenus:2011wh}
T.T.~Jouttenus, I.W.~Stewart, F.J.~Tackmann and W.J.~Waalewijn, \emph{{The Soft Function for Exclusive N-Jet Production at Hadron Colliders}}, \href{https://doi.org/10.1103/PhysRevD.83.114030}{\emph{Phys. Rev. D} {\bfseries 83} (2011) 114030} [\href{https://arxiv.org/abs/1102.4344}{{\ttfamily 1102.4344}}].

\bibitem{Gaunt:2015pea}
J.~Gaunt, M.~Stahlhofen, F.J.~Tackmann and J.R.~Walsh, \emph{{N-jettiness Subtractions for NNLO QCD Calculations}}, \href{https://doi.org/10.1007/JHEP09(2015)058}{\emph{JHEP} {\bfseries 09} (2015) 058} [\href{https://arxiv.org/abs/1505.04794}{{\ttfamily 1505.04794}}].

\bibitem{Cacciari:2015jma}
M.~Cacciari, F.A.~Dreyer, A.~Karlberg, G.P.~Salam and G.~Zanderighi, \emph{{Fully Differential Vector-Boson-Fusion Higgs Production at Next-to-Next-to-Leading Order}}, \href{https://doi.org/10.1103/PhysRevLett.115.082002}{\emph{Phys. Rev. Lett.} {\bfseries 115} (2015) 082002} [\href{https://arxiv.org/abs/1506.02660}{{\ttfamily 1506.02660}}].

\bibitem{Magnea:2018hab}
L.~Magnea, E.~Maina, G.~Pelliccioli, C.~Signorile-Signorile, P.~Torrielli and S.~Uccirati, \emph{{Local analytic sector subtraction at NNLO}}, \href{https://doi.org/10.1007/JHEP12(2018)107}{\emph{JHEP} {\bfseries 12} (2018) 107} [\href{https://arxiv.org/abs/1806.09570}{{\ttfamily 1806.09570}}].

\bibitem{Bertolotti:2022aih}
G.~Bertolotti, L.~Magnea, G.~Pelliccioli, A.~Ratti, C.~Signorile-Signorile, P.~Torrielli et~al., \emph{{NNLO subtraction for any massless final state: a complete analytic expression}}, \href{https://doi.org/10.1007/JHEP07(2023)140}{\emph{JHEP} {\bfseries 07} (2023) 140} [\href{https://arxiv.org/abs/2212.11190}{{\ttfamily 2212.11190}}].

\bibitem{Stewart:2010tn}
I.W.~Stewart, F.J.~Tackmann and W.J.~Waalewijn, \emph{{N-Jettiness: An Inclusive Event Shape to Veto Jets}}, \href{https://doi.org/10.1103/PhysRevLett.105.092002}{\emph{Phys. Rev. Lett.} {\bfseries 105} (2010) 092002} [\href{https://arxiv.org/abs/1004.2489}{{\ttfamily 1004.2489}}].

\bibitem{Buonocore:2023rdw}
L.~Buonocore, M.~Grazzini, J.~Haag, L.~Rottoli and C.~Savoini, \emph{{Exploring slicing variables for jet processes}}, \href{https://doi.org/10.1007/JHEP12(2023)193}{\emph{JHEP} {\bfseries 12} (2023) 193} [\href{https://arxiv.org/abs/2307.11570}{{\ttfamily 2307.11570}}].

\bibitem{Fu:2024fgj}
R.-J.~Fu, R.~Rahn, D.Y.~Shao, W.J.~Waalewijn and B.~Wu, \emph{{$q_T$-slicing with multiple jets}},  \href{https://arxiv.org/abs/2412.05358}{{\ttfamily 2412.05358}}.

\bibitem{Boughezal:2015dva}
R.~Boughezal, C.~Focke, X.~Liu and F.~Petriello, \emph{{$W$-boson production in association with a jet at next-to-next-to-leading order in perturbative QCD}}, \href{https://doi.org/10.1103/PhysRevLett.115.062002}{\emph{Phys. Rev. Lett.} {\bfseries 115} (2015) 062002} [\href{https://arxiv.org/abs/1504.02131}{{\ttfamily 1504.02131}}].

\bibitem{Devoto:2023rpv}
F.~Devoto, K.~Melnikov, R.~R\"ontsch, C.~Signorile-Signorile and D.M.~Tagliabue, \emph{{A fresh look at the nested soft-collinear subtraction scheme: NNLO QCD corrections to N-gluon final states in $ q\overline{q} $ annihilation}}, \href{https://doi.org/10.1007/JHEP02(2024)016}{\emph{JHEP} {\bfseries 02} (2024) 016} [\href{https://arxiv.org/abs/2310.17598}{{\ttfamily 2310.17598}}].

\bibitem{Devoto:2025kin}
F.~Devoto, K.~Melnikov, R.~R{\"o}ntsch, C.~Signorile-Signorile, D.M.~Tagliabue and M.~Tresoldi, \emph{{Towards a general subtraction formula for NNLO QCD corrections to processes at hadron colliders: final states with quarks and gluons}}, \href{https://doi.org/10.1007/JHEP08(2025)122}{\emph{JHEP} {\bfseries 08} (2025) 122} [\href{https://arxiv.org/abs/2503.15251}{{\ttfamily 2503.15251}}].

\bibitem{Devoto:2025jql}
F.~Devoto, K.~Melnikov, R.~R{\"o}ntsch, C.~Signorile-Signorile, D.M.~Tagliabue and M.~Tresoldi, \emph{{Integrated subtraction terms and finite remainders for arbitrary processes with massless partons at colliders in the nested soft-collinear subtraction scheme}},  \href{https://arxiv.org/abs/2509.08594}{{\ttfamily 2509.08594}}.

\bibitem{Fox:2024bfp}
E.~Fox, N.~Glover and M.~Marcoli, \emph{{Generalised antenna functions for higher-order calculations}}, \href{https://doi.org/10.1007/JHEP12(2024)225}{\emph{JHEP} {\bfseries 12} (2024) 225} [\href{https://arxiv.org/abs/2410.12904}{{\ttfamily 2410.12904}}].

\bibitem{Gehrmann:2023dxm}
T.~Gehrmann, E.W.N.~Glover and M.~Marcoli, \emph{{The colourful antenna subtraction method}}, \href{https://doi.org/10.1007/JHEP03(2024)114}{\emph{JHEP} {\bfseries 03} (2024) 114} [\href{https://arxiv.org/abs/2310.19757}{{\ttfamily 2310.19757}}].

\bibitem{vanBeekveld:2022zhl}
M.~van Beekveld, S.~Ferrario~Ravasio, G.P.~Salam, A.~Soto-Ontoso, G.~Soyez and R.~Verheyen, \emph{{PanScales parton showers for hadron collisions: formulation and fixed-order studies}}, \href{https://doi.org/10.1007/JHEP11(2022)019}{\emph{JHEP} {\bfseries 11} (2022) 019} [\href{https://arxiv.org/abs/2205.02237}{{\ttfamily 2205.02237}}].

\bibitem{vanBeekveld:2025lpz}
M.~van Beekveld, S.~Ferrario~Ravasio, J.~Helliwell, A.~Karlberg, G.P.~Salam, L.~Scyboz et~al., \emph{{Logarithmically-accurate and positive-definite NLO shower matching}},  \href{https://arxiv.org/abs/2504.05377}{{\ttfamily 2504.05377}}.

\bibitem{FerrarioRavasio:2023kyg}
S.~Ferrario~Ravasio, K.~Hamilton, A.~Karlberg, G.P.~Salam, L.~Scyboz and G.~Soyez, \emph{{Parton Showering with Higher Logarithmic Accuracy for Soft Emissions}}, \href{https://doi.org/10.1103/PhysRevLett.131.161906}{\emph{Phys. Rev. Lett.} {\bfseries 131} (2023) 161906} [\href{https://arxiv.org/abs/2307.11142}{{\ttfamily 2307.11142}}].

\bibitem{Forshaw:2025bmo}
J.R.~Forshaw, S.~Pl{\"a}tzer and F.T.~Gonz{\'a}lez, \emph{{Exact colour evolution for jet observables}},  \href{https://arxiv.org/abs/2502.12133}{{\ttfamily 2502.12133}}.

\bibitem{Forshaw:2025fif}
J.R.~Forshaw, S.~Pl{\"a}tzer and F.T.~Gonz{\'a}lez, \emph{{Fully Differential Soft Gluon Evolution at the Amplitude Level}},  \href{https://arxiv.org/abs/2505.13183}{{\ttfamily 2505.13183}}.

\bibitem{Altmann:2025yip}
J.~Altmann, H.T.~Li, L.~Scyboz and P.~Skands, \emph{{Sudakov evolution without unitarity}}, \href{https://doi.org/10.1140/epjc/s10052-025-14576-1}{\emph{Eur. Phys. J. C} {\bfseries 85} (2025) 840} [\href{https://arxiv.org/abs/2507.00111}{{\ttfamily 2507.00111}}].

\bibitem{El-Menoufi:2024sys}
B.K.~El-Menoufi, C.T.~Preuss, L.~Scyboz and P.~Skands, \emph{{Matching Z $\to$ Hadrons at NNLO with Sector Showers}},  \href{https://arxiv.org/abs/2412.14242}{{\ttfamily 2412.14242}}.

\bibitem{Herren:2022jej}
F.~Herren, S.~H{\"o}che, F.~Krauss, D.~Reichelt and M.~Schoenherr, \emph{{A new approach to color-coherent parton evolution}}, \href{https://doi.org/10.1007/JHEP10(2023)091}{\emph{JHEP} {\bfseries 10} (2023) 091} [\href{https://arxiv.org/abs/2208.06057}{{\ttfamily 2208.06057}}].

\bibitem{Skands:2023lkp}
P.~Skands and C.T.~Preuss, \emph{{NNLO Matrix-Element Corrections in VINCIA}}, \href{https://doi.org/10.22323/1.432.0067}{\emph{PoS} {\bfseries RADCOR2023} (2024) 067} [\href{https://arxiv.org/abs/2310.18671}{{\ttfamily 2310.18671}}].

\bibitem{Monni:2019whf}
P.F.~Monni, P.~Nason, E.~Re, M.~Wiesemann and G.~Zanderighi, \emph{{MiNNLO$_{PS}$: a new method to match NNLO QCD to parton showers}}, \href{https://doi.org/10.1007/JHEP05(2020)143}{\emph{JHEP} {\bfseries 05} (2020) 143} [\href{https://arxiv.org/abs/1908.06987}{{\ttfamily 1908.06987}}].

\bibitem{Alioli:2021qbf}
S.~Alioli, C.W.~Bauer, A.~Broggio, A.~Gavardi, S.~Kallweit, M.A.~Lim et~al., \emph{{Matching NNLO predictions to parton showers using N3LL color-singlet transverse momentum resummation in geneva}}, \href{https://doi.org/10.1103/PhysRevD.104.094020}{\emph{Phys. Rev. D} {\bfseries 104} (2021) 094020} [\href{https://arxiv.org/abs/2102.08390}{{\ttfamily 2102.08390}}].

\bibitem{Alioli:2025hpa}
S.~Alioli, G.~Billis, A.~Broggio and G.~Stagnitto, \emph{{NNLO predictions with nonlocal subtractions and fiducial power corrections in GENEVA}},  \href{https://arxiv.org/abs/2504.11357}{{\ttfamily 2504.11357}}.

\bibitem{Hoang:2024nqi}
A.H.~Hoang, O.L.~Jin, S.~Pl{\"a}tzer and D.~Samitz, \emph{{Matching hadronization and perturbative evolution: the cluster model in light of infrared shower cutoff dependence}}, \href{https://doi.org/10.1007/JHEP07(2025)005}{\emph{JHEP} {\bfseries 07} (2025) 005} [\href{https://arxiv.org/abs/2404.09856}{{\ttfamily 2404.09856}}].

\bibitem{Gieseke:2025mcy}
S.~Gieseke, S.~Kiebacher, S.~Pl{\"a}tzer and J.~Priedigkeit, \emph{{Phenomenological constraints of the building blocks of the cluster hadronization model}}, \href{https://doi.org/10.1140/epjc/s10052-025-14511-4}{\emph{Eur. Phys. J. C} {\bfseries 85} (2025) 796} [\href{https://arxiv.org/abs/2505.14542}{{\ttfamily 2505.14542}}].

\bibitem{Cieri:2019tfv}
L.~Cieri, C.~Oleari and M.~Rocco, \emph{{Higher-order power corrections in a transverse-momentum cut for colour-singlet production at NLO}}, \href{https://doi.org/10.1140/epjc/s10052-019-7361-8}{\emph{Eur. Phys. J. C} {\bfseries 79} (2019) 852} [\href{https://arxiv.org/abs/1906.09044}{{\ttfamily 1906.09044}}].

\bibitem{Boughezal:2016zws}
R.~Boughezal, X.~Liu and F.~Petriello, \emph{{Power Corrections in the N-jettiness Subtraction Scheme}}, \href{https://doi.org/10.1007/JHEP03(2017)160}{\emph{JHEP} {\bfseries 03} (2017) 160} [\href{https://arxiv.org/abs/1612.02911}{{\ttfamily 1612.02911}}].

\bibitem{DelDuca:2017twk}
V.~Del~Duca, E.~Laenen, L.~Magnea, L.~Vernazza and C.D.~White, \emph{{Universality of next-to-leading power threshold effects for colourless final states in hadronic collisions}}, \href{https://doi.org/10.1007/JHEP11(2017)057}{\emph{JHEP} {\bfseries 11} (2017) 057} [\href{https://arxiv.org/abs/1706.04018}{{\ttfamily 1706.04018}}].

\bibitem{Boughezal:2018mvf}
R.~Boughezal, A.~Isgr\`o and F.~Petriello, \emph{{Next-to-leading-logarithmic power corrections for $N$-jettiness subtraction in color-singlet production}}, \href{https://doi.org/10.1103/PhysRevD.97.076006}{\emph{Phys. Rev. D} {\bfseries 97} (2018) 076006} [\href{https://arxiv.org/abs/1802.00456}{{\ttfamily 1802.00456}}].

\bibitem{Moult:2016fqy}
I.~Moult, L.~Rothen, I.W.~Stewart, F.J.~Tackmann and H.X.~Zhu, \emph{{Subleading Power Corrections for N-Jettiness Subtractions}}, \href{https://doi.org/10.1103/PhysRevD.95.074023}{\emph{Phys. Rev. D} {\bfseries 95} (2017) 074023} [\href{https://arxiv.org/abs/1612.00450}{{\ttfamily 1612.00450}}].

\bibitem{Moult:2017jsg}
I.~Moult, L.~Rothen, I.W.~Stewart, F.J.~Tackmann and H.X.~Zhu, \emph{{N -jettiness subtractions for $gg\to H$ at subleading power}}, \href{https://doi.org/10.1103/PhysRevD.97.014013}{\emph{Phys. Rev. D} {\bfseries 97} (2018) 014013} [\href{https://arxiv.org/abs/1710.03227}{{\ttfamily 1710.03227}}].

\bibitem{Ebert:2018lzn}
M.A.~Ebert, I.~Moult, I.W.~Stewart, F.J.~Tackmann, G.~Vita and H.X.~Zhu, \emph{{Power Corrections for N-Jettiness Subtractions at ${\cal O}(\alpha_s)$}}, \href{https://doi.org/10.1007/JHEP12(2018)084}{\emph{JHEP} {\bfseries 12} (2018) 084} [\href{https://arxiv.org/abs/1807.10764}{{\ttfamily 1807.10764}}].

\bibitem{Boughezal:2019ggi}
R.~Boughezal, A.~Isgr\`o and F.~Petriello, \emph{{Next-to-leading power corrections to $V+1$ jet production in $N$-jettiness subtraction}}, \href{https://doi.org/10.1103/PhysRevD.101.016005}{\emph{Phys. Rev. D} {\bfseries 101} (2020) 016005} [\href{https://arxiv.org/abs/1907.12213}{{\ttfamily 1907.12213}}].

\bibitem{vanBeekveld:2019prq}
M.~van Beekveld, W.~Beenakker, E.~Laenen and C.D.~White, \emph{{Next-to-leading power threshold effects for inclusive and exclusive processes with final state jets}}, \href{https://doi.org/10.1007/JHEP03(2020)106}{\emph{JHEP} {\bfseries 03} (2020) 106} [\href{https://arxiv.org/abs/1905.08741}{{\ttfamily 1905.08741}}].

\bibitem{Oleari:2020wvt}
C.~Oleari and M.~Rocco, \emph{{Power corrections in a transverse-momentum cut for vector-boson production at NNLO: the $qg$-initiated real-virtual contribution}}, \href{https://doi.org/10.1140/epjc/s10052-021-08878-3}{\emph{Eur. Phys. J. C} {\bfseries 81} (2021) 183} [\href{https://arxiv.org/abs/2012.10538}{{\ttfamily 2012.10538}}].

\bibitem{Vita:2024ypr}
G.~Vita, \emph{{N$^{3}$LO power corrections for 0-jettiness subtractions with fiducial cuts}}, \href{https://doi.org/10.1007/JHEP07(2024)241}{\emph{JHEP} {\bfseries 07} (2024) 241} [\href{https://arxiv.org/abs/2401.03017}{{\ttfamily 2401.03017}}].

\bibitem{Pal:2023vec}
S.~Pal and S.~Seth, \emph{{On Higgs+jet production at next-to-leading power accuracy}}, \href{https://doi.org/10.1103/PhysRevD.109.114018}{\emph{Phys. Rev. D} {\bfseries 109} (2024) 114018} [\href{https://arxiv.org/abs/2309.08343}{{\ttfamily 2309.08343}}].

\bibitem{Pal:2024eyr}
S.~Pal and S.~Seth, \emph{{Soft quark effects on H+jet production at NLP accuracy}}, \href{https://doi.org/10.1016/j.physletb.2024.139179}{\emph{Phys. Lett. B} {\bfseries 860} (2025) 139179} [\href{https://arxiv.org/abs/2405.06444}{{\ttfamily 2405.06444}}].

\bibitem{Beneke:2018gvs}
M.~Beneke, A.~Broggio, M.~Garny, S.~Jaskiewicz, R.~Szafron, L.~Vernazza et~al., \emph{{Leading-logarithmic threshold resummation of the Drell-Yan process at next-to-leading power}}, \href{https://doi.org/10.1007/JHEP03(2019)043}{\emph{JHEP} {\bfseries 03} (2019) 043} [\href{https://arxiv.org/abs/1809.10631}{{\ttfamily 1809.10631}}].

\bibitem{Beneke:2019oqx}
M.~Beneke, A.~Broggio, S.~Jaskiewicz and L.~Vernazza, \emph{{Threshold factorization of the Drell-Yan process at next-to-leading power}}, \href{https://doi.org/10.1007/JHEP07(2020)078}{\emph{JHEP} {\bfseries 07} (2020) 078} [\href{https://arxiv.org/abs/1912.01585}{{\ttfamily 1912.01585}}].

\bibitem{Bonocore:2015esa}
D.~Bonocore, E.~Laenen, L.~Magnea, S.~Melville, L.~Vernazza and C.D.~White, \emph{{A factorization approach to next-to-leading-power threshold logarithms}}, \href{https://doi.org/10.1007/JHEP06(2015)008}{\emph{JHEP} {\bfseries 06} (2015) 008} [\href{https://arxiv.org/abs/1503.05156}{{\ttfamily 1503.05156}}].

\bibitem{Bonocore:2016awd}
D.~Bonocore, E.~Laenen, L.~Magnea, L.~Vernazza and C.D.~White, \emph{{Non-abelian factorisation for next-to-leading-power threshold logarithms}}, \href{https://doi.org/10.1007/JHEP12(2016)121}{\emph{JHEP} {\bfseries 12} (2016) 121} [\href{https://arxiv.org/abs/1610.06842}{{\ttfamily 1610.06842}}].

\bibitem{Broggio:2021fnr}
A.~Broggio, S.~Jaskiewicz and L.~Vernazza, \emph{{Next-to-leading power two-loop soft functions for the Drell-Yan process at threshold}}, \href{https://doi.org/10.1007/JHEP10(2021)061}{\emph{JHEP} {\bfseries 10} (2021) 061} [\href{https://arxiv.org/abs/2107.07353}{{\ttfamily 2107.07353}}].

\bibitem{Broggio:2023pbu}
A.~Broggio, S.~Jaskiewicz and L.~Vernazza, \emph{{Threshold factorization of the Drell-Yan quark-gluon channel and two-loop soft function at next-to-leading power}}, \href{https://doi.org/10.1007/JHEP12(2023)028}{\emph{JHEP} {\bfseries 12} (2023) 028} [\href{https://arxiv.org/abs/2306.06037}{{\ttfamily 2306.06037}}].

\bibitem{Ebert:2018gsn}
M.A.~Ebert, I.~Moult, I.W.~Stewart, F.J.~Tackmann, G.~Vita and H.X.~Zhu, \emph{{Subleading power rapidity divergences and power corrections for q$_{T}$}}, \href{https://doi.org/10.1007/JHEP04(2019)123}{\emph{JHEP} {\bfseries 04} (2019) 123} [\href{https://arxiv.org/abs/1812.08189}{{\ttfamily 1812.08189}}].

\bibitem{Laenen:2020nrt}
E.~Laenen, J.~Sinninghe~Damst\'e, L.~Vernazza, W.~Waalewijn and L.~Zoppi, \emph{{Towards all-order factorization of QED amplitudes at next-to-leading power}}, \href{https://doi.org/10.1103/PhysRevD.103.034022}{\emph{Phys. Rev. D} {\bfseries 103} (2021) 034022} [\href{https://arxiv.org/abs/2008.01736}{{\ttfamily 2008.01736}}].

\bibitem{Pal:2025ffp}
S.~Pal and S.~Seth, \emph{{Universality at next-to-leading power for jet associated processes}},  \href{https://arxiv.org/abs/2505.01340}{{\ttfamily 2505.01340}}.

\bibitem{Czakon:2023tld}
M.~Czakon, F.~Eschment and T.~Schellenberger, \emph{{Subleading effects in soft-gluon emission at one-loop in massless QCD}}, \href{https://doi.org/10.1007/JHEP12(2023)126}{\emph{JHEP} {\bfseries 12} (2023) 126} [\href{https://arxiv.org/abs/2307.02286}{{\ttfamily 2307.02286}}].

\bibitem{Agarwal:2025dvo}
P.~Agarwal, K.~Melnikov, I.~Pedron and P.~Pfohl, \emph{{Power corrections to the production of a color-singlet final state in hadron collisions in the N-jettiness slicing scheme at NLO QCD}},  \href{https://arxiv.org/abs/2502.09327}{{\ttfamily 2502.09327}}.

\bibitem{Frixione:1998jh}
S.~Frixione, \emph{{Isolated photons in perturbative QCD}}, \href{https://doi.org/10.1016/S0370-2693(98)00454-7}{\emph{Phys. Lett. B} {\bfseries 429} (1998) 369} [\href{https://arxiv.org/abs/hep-ph/9801442}{{\ttfamily hep-ph/9801442}}].

\bibitem{Ellis:1993tq}
S.D.~Ellis and D.E.~Soper, \emph{{Successive combination jet algorithm for hadron collisions}}, \href{https://doi.org/10.1103/PhysRevD.48.3160}{\emph{Phys. Rev. D} {\bfseries 48} (1993) 3160} [\href{https://arxiv.org/abs/hep-ph/9305266}{{\ttfamily hep-ph/9305266}}].

\bibitem{Catani:1993hr}
S.~Catani, Y.L.~Dokshitzer, M.H.~Seymour and B.R.~Webber, \emph{{Longitudinally invariant $K_t$ clustering algorithms for hadron hadron collisions}}, \href{https://doi.org/10.1016/0550-3213(93)90166-M}{\emph{Nucl. Phys. B} {\bfseries 406} (1993) 187}.

\bibitem{Ebert:2019zkb}
M.A.~Ebert and F.J.~Tackmann, \emph{{Impact of isolation and fiducial cuts on q$_{T}$ and N-jettiness subtractions}}, \href{https://doi.org/10.1007/JHEP03(2020)158}{\emph{JHEP} {\bfseries 03} (2020) 158} [\href{https://arxiv.org/abs/1911.08486}{{\ttfamily 1911.08486}}].

\bibitem{Salam:2021tbm}
G.P.~Salam and E.~Slade, \emph{{Cuts for two-body decays at colliders}}, \href{https://doi.org/10.1007/JHEP11(2021)220}{\emph{JHEP} {\bfseries 11} (2021) 220} [\href{https://arxiv.org/abs/2106.08329}{{\ttfamily 2106.08329}}].

\bibitem{Ebert:2020dfc}
M.A.~Ebert, J.K.L.~Michel, I.W.~Stewart and F.J.~Tackmann, \emph{{Drell-Yan $q_{T}$ resummation of fiducial power corrections at N$^{3}$LL}}, \href{https://doi.org/10.1007/JHEP04(2021)102}{\emph{JHEP} {\bfseries 04} (2021) 102} [\href{https://arxiv.org/abs/2006.11382}{{\ttfamily 2006.11382}}].

\bibitem{Campbell:2024hjq}
J.~Campbell, T.~Neumann and G.~Vita, \emph{{Projection-to-Born-improved subtractions at NNLO}}, \href{https://doi.org/10.1007/JHEP05(2025)172}{\emph{JHEP} {\bfseries 05} (2025) 172} [\href{https://arxiv.org/abs/2408.05265}{{\ttfamily 2408.05265}}].

\bibitem{Burnett:1967km}
T.H.~Burnett and N.M.~Kroll, \emph{{Extension of the low soft photon theorem}}, \href{https://doi.org/10.1103/PhysRevLett.20.86}{\emph{Phys. Rev. Lett.} {\bfseries 20} (1968) 86}.

\bibitem{Low:1958sn}
F.E.~Low, \emph{{Bremsstrahlung of very low-energy quanta in elementary particle collisions}}, \href{https://doi.org/10.1103/PhysRev.110.974}{\emph{Phys. Rev.} {\bfseries 110} (1958) 974}.

\bibitem{DelDuca:2019ctm}
V.~Del~Duca, N.~Deutschmann and S.~Lionetti, \emph{{Momentum mappings for subtractions at higher orders in QCD}}, \href{https://doi.org/10.1007/JHEP12(2019)129}{\emph{JHEP} {\bfseries 12} (2019) 129} [\href{https://arxiv.org/abs/1910.01024}{{\ttfamily 1910.01024}}].

\bibitem{Catani:2000ef}
S.~Catani, S.~Dittmaier and Z.~Trocsanyi, \emph{{One loop singular behavior of QCD and SUSY QCD amplitudes with massive partons}}, \href{https://doi.org/10.1016/S0370-2693(01)00065-X}{\emph{Phys. Lett. B} {\bfseries 500} (2001) 149} [\href{https://arxiv.org/abs/hep-ph/0011222}{{\ttfamily hep-ph/0011222}}].

\end{thebibliography}\endgroup

\end{document}